\documentclass[journal=jctcce,manuscript=article]{achemso}

\usepackage[version=3]{mhchem} 
\usepackage{braket}
\usepackage{mathtools}
\usepackage{algorithm}
\usepackage{algorithmic}
\usepackage{geometry}
\usepackage{amsmath}
\usepackage{amsthm}
\usepackage{graphicx}
\usepackage{fancyvrb}
\usepackage{mciteplus}
\usepackage{indentfirst}
\usepackage[dvipsnames]{xcolor}
\usepackage{xurl}
\usepackage{hyperref}

\newcommand{\pfn}{\textbf{P}_{\rm Fn}(t)}
\newcommand{\pe}{\textbf{P}_{\rm e}(t)}
\newcommand{\pn}{\textbf{P}_{\rm n}(t)}
\newcommand{\pf}{\textbf{P}_{\rm F}(t)}
\newcommand{\smax}{S_{\rm max}}
\newcommand{\s}{S(t)}
\newcommand{\mandelq}{Q_{k, \lambda}(t)}
\newcommand{\mandelqx}{Q_{k, x}(t)}
\newcommand{\nav}{n_{k, x}(t)}
\newcommand{\varn}{{\rm var}\left[n_{k, x}\right](t)}
\newcommand{\navl}{n_{k, \lambda}(t)}
\newcommand{\varnl}{{\rm var}\left[n_{k, \lambda}\right](t)}

\newcommand{\varxx}{{\rm var}\left[X_{k,x}\right](t)}
\newcommand{\varyx}{{\rm var}\left[Y_{k,x}\right](t)}
\newcommand{\varxl}{{\rm var}\left[X_{k,\lambda}\right](t)}
\newcommand{\varyl}{{\rm var}\left[Y_{k,\lambda}\right](t)}
\newcommand{\te}{{T}_{\rm E}}
\newcommand{\tc}{{T}_{\rm c}}
\newcommand{\tr}{{T}_{\rm R}}
\newcommand{\xl}{\hat{X}_{k, \lambda}}

\newcommand{\yl}{\hat{Y}_{k, \lambda}}

\author{Millan F. Welman}
\affiliation[Princeton University]
{Department of Chemistry, Princeton University, Princeton, NJ 08544}
\author{Sharon Hammes-Schiffer}
\email{shs566@princeton.edu}

\affiliation[Princeton University]
{Department of Chemistry, Princeton University, Princeton, NJ 08544}

\title{Photon Statistics and Light Squeezing Under Vibrational Strong Coupling from Real-Time Nuclear--Electronic Orbital Theory}


\begin{document}

\begin{abstract}
    Molecular polaritons have attracted great interest in recent years as a promising avenue for developing optical control of chemical processes. A major open question in this field is the role of quantum optical effects in understanding the molecular dynamics observed in cavities. In this work, we use our recently developed mean-field quantum and full-quantum real-time nuclear--electronic orbital time-dependent density functional theory methods to simulate photon statistics for a cavity mode strongly coupled to a molecular vibrational transition. Our comparative use of these two methods allows us to isolate the contributions of light--matter entanglement to the photon statistics. We consider three initial conditions of the cavity mode: a coherent state, a squeezed coherent state, and a single photon Fock state. These conditions span Poissonian, super-Poissonian, and sub-Poissonian photon statistics. Our full-quantum dynamics initialized with a coherent state show considerable super-Poissonian fluctuations of the photon statistics and light-squeezing of the cavity photons by the vibrational mode. These dynamics are particularly notable considering the quasi-classical properties of the coherent state. The squeezed coherent and Fock state initial conditions also demonstrate noteworthy fluctuations in the photon statistics. All three initial conditions reveal dynamics in the full-quantum photon statistics that differ considerably from the statistics obtained with the mean-field quantum method. Our results demonstrate the importance of considering quantum optical effects for molecules under strong coupling in certain situations. 
\end{abstract}
\section{Introduction}

Molecular polaritons, quasiparticles formed by the strong coupling of light and matter degrees of freedom, have gained increased recognition in recent years as a promising avenue for achieving optical control over chemical processes \cite{mandal_theoretical_2023}. 
In vibrational strong coupling (VSC), a cavity mode (i.e., a standing electromagnetic wave in an optical cavity) is strongly coupled to a vibrational transition\cite{long_coherent_2015, george_liquid-phase_2015, wright_rovibrational_2023}, whereas in electronic strong coupling (ESC), the cavity mode is strongly coupled to an electronic transition.\cite{deng_exciton-polariton_2010, keeling_boseeinstein_2020}. Experimental work performed under either VSC or ESC has shown the possibility of modifying chemical processes ranging from energy transfer and exciton transport processes\cite{zhong_energy_2017, xiang_intermolecular_2020, garcia-vidal_manipulating_2021, georgiou_ultralongrange_2021, kertzscher_tunable_2024, lidzey_strong_1998, feist_extraordinary_2015, schachenmayer_cavity-enhanced_2015, coles_polariton-mediated_2014, kasprzak_boseeinstein_2006} to reaction rates \cite{ahn_modification_2023, thomas_ground_2020, lather_cavity_2019} and branching ratios \cite{thomas_tilting_2019}.

From a theoretical point of view, one major question for studying polaritons is determining whether to treat the cavity modes with classical or quantum optics\cite{schwennicke_when_2025}. 
Either choice is valid because the coherent exchange of energy between matter and cavity mode degrees of freedom that constitutes the formation of a polaritonic state is predicted by both classical and quantum electrodynamics.
Although canonical quantum optical models \cite{jaynes_comparison_1963, tavis_exact_1968, tavis_approximate_1969} remain central to modern theoretical studies of polaritons, it is well-established that the light--matter hybridization necessary for polariton formation may still occur if the cavity modes are treated as classical standing waves \cite{hoffmann_effect_2020, li_cavity_2020, garner_simulation_2025, fetherolf_nuclearelectronic_2026}. 
It is indeed known that classical transfer matrix methods can accurately reproduce experimentally measured polaritonic spectra \cite{zhu_vacuum_1990, mckillop_cavity-enhanced_2025, wright_rovibrational_2023, chen_exploring_2024}. 
This result has been supported by recent theoretical work in the many-molecule collective regime, \cite{yuen-zhou_linear_2024} where most experiments on cavity-modified chemical dynamics are performed. 
From the perspective of simulation, this result is appealing because of the relatively low cost of simulating classical optics. 
This low cost makes classical treatments of the cavity modes a practical route for studying the collective regime, a direction that has been explored in recent years to yield considerable insight into non-reactive dynamics under VSC \cite{li_collective_2021, li_energy-efficient_2022, li_polariton_2022, li_qmmm_2023, li_quantum_2022, li_mesoscale_2024, li_fdtd_2025, ji_selective_2025, bocanegra_vargas_polariton-induced_2025}. 

In spite of this progress, important questions remain to be examined by considering a quantized cavity mode.
Quantized cavity modes are necessary for the description of important phenomena such as spontaneous emission and light--matter entanglement. 
Although simulating the first-principles dynamics of quantized cavity modes coupled to many molecules is not generally a computationally tractable problem, the simulation of a single molecule interacting with a few quantized cavity modes, as might occur experimentally in a plasmonic nanocavity \cite{pelton_strong_2019, li_molecular_2022, xiang_molecular_2024}, is within reach. 
Furthermore, recent theoretical work suggests that modified dynamics \cite{flick_atoms_2017, ruggenthaler_quantum-electrodynamical_2014, flick_kohnsham_2015, fiechter_how_2023}, including modified spectral features \cite{simko_twin_2025, koner_hidden_2025} and reaction rates \cite{ke_quantum_2024}, may manifest under strong coupling to a quantized cavity mode as compared to a classical cavity mode. 
Therefore, it is valuable to understand the results of quantum mechanical treatments of the cavity mode when considering the tractable case of single-molecule polaritons and how these results might differ from those obtained with a classical cavity mode. 

To this end, we have recently introduced a hierarchy \cite{li_semiclassical_2022, welman_light-matter_2025} of first-principles dynamical simulation methods for molecular polaritons under VSC or ESC. 
For the case of VSC, the molecular degrees of freedom are treated with real-time nuclear--electronic orbital time-dependent density-functional theory (hereafter RT-NEO)\cite{zhao_real-time_2020}. 
RT-NEO theory generalizes conventional electronic RT-TDDFT within the NEO framework. In the NEO framework,\cite{webb_multiconfigurational_2002, hammes-schiffer_nuclearelectronic_2021} specified light nuclei, usually protons, are treated quantum mechanically on the same footing as the electronic degrees of freedom. 
The RT-NEO method incorporates zero-point energy, tunneling, proton delocalization, and other nuclear quantum effects in its propagation of molecular quantum dynamics. 
Coupling RT-NEO to the real-time propagation of one or more cavity modes therefore provides a route to simulating polariton dynamics under VSC in a manner that includes nuclear quantum effects for the vibrational degrees of freedom that are strongly coupled to the cavity.
In this work, we will use the specific flavor of RT-NEO known as electronic Born-Oppenheimer RT-NEO (BO-RT-NEO), which we have successfully applied in previous work \cite{li_electronic_2023, welman_light-matter_2025} to the simulation of real-time polariton dynamics under VSC. For notational simplicity, we will refer to BO-RT-NEO as simply RT-NEO in this work.

Our hierarchy has three methods, two of which will be used in this manuscript. 
The full-quantum RT-NEO (fq-RT-NEO) method propagates a joint light--matter density matrix describing both the cavity photons and the quantum nuclear system to which they are strongly coupled. 
This approach includes the effects of light--matter entanglement, vacuum fluctuations, and hybridization between light and matter degrees of freedom. 
In previous work, we used the fq-RT-NEO method to study the dynamics of the von Neumann entanglement entropy between light and matter degrees of freedom \cite{welman_light-matter_2025}, as well as dynamics where the cavity mode is initialized in the single-excitation manifold instead of as a coherent state \cite{welman_initialization_2026}.
In contrast to the fq-RT-NEO method, the mean-field-quantum RT-NEO (mfq-RT-NEO) method self-consistently propagates separate equations of motion for the density matrices, one for the cavity photons and the other for the quantum nuclear degrees of freedom. 
This method is equivalent to neglecting light--matter entanglement while still considering vacuum fluctuations and light--matter hybridization. 
By comparing the results of the fq-RT-NEO and mfq-RT-NEO methods, we can identify the contributions of light--matter entanglement to the dynamics of strongly coupled light--matter systems. An advantage of this strategy is that the molecular degrees of freedom are treated with a realistic, first-principles approach.

In this work, we will use the fq- and mfq-RT-NEO methods to compute cavity mode photon statistics. 
Analysis of the statistics and correlations of the photon number has been shown to be a useful tool for experimental and theoretical analysis of light--matter interactions, both inside \cite{rundquist_nonclassical_2014, saez-blazquez_enhancing_2017, flick_kohnsham_2015, stevens_third-order_2014, hoffmann_capturing_2019, baryshev_engineering_2022, love_intrinsic_2008, saez-blazquez_photon_2018, putintsev_photon_2024, whittaker_polariton_2017, hofmann_sub-poissonian_2013, kam_super-poissonian_2026, tasci_super-poissonian_2025} and outside \cite{cao_correlations_2006, vester_photon_2015, amgar_higher-order_2019} an optical cavity. 
Our work will center on the Mandel Q Parameter \cite{Mandel:79}, which provides a natural metric for measuring deviations of the cavity field from the behavior associated with a quasi-classical (coherent) field. 
For each method, we will compare the dynamics of this parameter for three different cavity mode initial states, each of which is chosen to be representative of a given regime of photon statistics. 
We will also compare the dynamics of the Mandel Q Parameter to those of other cavity mode observables obtained from the fq- and mfq-RT-NEO methods. 
By carrying out this analysis using our polaritonic RT-NEO hierarchy, we can simulate cavity photon statistics for a cavity mode strongly coupled to a molecule treated with first-principles real-time electronic structure theory. 

The remainder of the paper is organized as follows. In section 2, we review the mfq- and fq-RT-NEO equations and outline our procedures for computing light--matter entanglement and photon statistics. Simulation details
for the HCN molecule under VSC are provided in section 3. The results of the mfq- and fq-RT-NEO calculations and supporting analysis are presented in section 4. Conclusions are given in section 5.  

\section{Theory}

In this section, we provide an overview of the fq-RT-NEO and mfq-RT-NEO equations and the properties of the von Neumann entropy and the Mandel Q Parameter. 
Except for the Mandel Q Parameter, all of these topics have been discussed extensively in our previous work \cite{welman_light-matter_2025, welman_initialization_2026}. 
We refer the interested reader to these papers for a more in-depth discussion of the derivation of the mfq- and fq-RT-NEO equations of motion and the properties of the von Neumann entropy beyond the summary provided here. 

In the fq-RT-NEO method, we propagate an equation of motion giving the time evolution of the joint density matrix $\pfn$ that describes the quantum nuclear subsystem and the cavity mode to which it is strongly coupled:
\begin{equation}
    i\hbar\frac{\partial}{\partial t}\textbf{P}_{\rm Fn}(t) = \left[\textbf{I}_{\rm F} \otimes \textbf{F}^{\rm{NEO}}_{\rm n}(t) + \textbf{H}_{{\rm F}} \otimes \textbf{I}_{\rm n} + \sum_{k,\lambda}\varepsilon_{k,\lambda} \textbf{q}_{k,\lambda}\otimes\left(\boldsymbol{\mu}_{{\rm n}, \lambda} - \mu^0_{{\rm n}, \lambda}\textbf{I}_{\rm n}\right), \,\textbf{P}_{\rm Fn}(t) \right]
    \label{rt neo nuc eom}
\end{equation}
This equation is written in the long-wavelength approximation and neglects the self-dipole interaction, an acceptable approximation here because we are ignoring the ultrastrong coupling regime and focusing on the strong coupling regime \cite{schafer_relevance_2020}. 
As the self-dipole term is entirely molecular in character, we do not expect it to significantly affect cavity mode observables, and it has indeed been successfully neglected in several previous model studies of cavity photon statistics \cite{carmichael_photon_1985, meystre_squeezed_1982, rundquist_nonclassical_2014, hoffmann_capturing_2019, saez-blazquez_enhancing_2017, saez-blazquez_photon_2018}. 
In these equations, the subscripts ``F" and``n" denote ``field" (cavity mode) and quantum nuclear degrees of freedom, respectively. 
In Eq. \ref{rt neo nuc eom}, $\textbf{I}_{\rm F}$ and $\textbf{I}_{\rm n}$ are the identity matrices in the cavity mode and quantum nuclear Hilbert spaces, respectively. 
The Kronecker product $\otimes$ expands operators in the individual cavity mode and quantum nuclear Hilbert spaces into the joint molecule--mode Hilbert space.
$\textbf{H}_{\rm F}$ is the matrix representation of the sum of bare cavity mode Hamiltonians given by
\begin{equation}
    \hat{H}_{\rm F} = 
    \sum_{k,\lambda}\left[\frac{1}{2}\hat{p}_{k,\lambda}^2 + \frac{1}{2}\omega_{k,\lambda}^2\hat{q}_{k,\lambda}^2\right]
    \label{free field hamiltonian}
\end{equation}
The sum appearing in Eqs. \ref{rt neo nuc eom} and \ref{free field hamiltonian} is over cavity modes, each of which is determined by a wavevector magnitude $k = |\boldsymbol{\rm k}|$ and polarization unit vector $\boldsymbol{\xi}_\lambda$ satisfying $\boldsymbol{\rm k}  \cdot \boldsymbol{\xi}_\lambda = 0$. 
Here, $\hat{q}_{k,\lambda}$, $\hat{p}_{k,\lambda}$, and $\omega_{k,\lambda}$ are the coordinate operator, momentum operator, and frequency for the mode $\{k, \lambda\}$. 
We also note that the wavevector magnitude satisfies $k = \omega_{k,\lambda}/c$, where $c$ is the speed of light.

In Eq. \ref{rt neo nuc eom}, the light--matter interaction potential is given by the Kronecker product of $\textbf{q}_{k,\lambda}$, the matrix representation of $\hat{q}_{k,\lambda}$, and $\boldsymbol{\mu}_{{\rm n}, \lambda}$, the matrix representation of the single-particle quantum nuclear dipole moment operator's component in the polarization direction $\lambda$. 
We subtract the component of the permanent quantum nuclear dipole moment in the polarization direction, $\mu^0_{{\rm n}, \lambda}$, to ensure that the cavity mode is not perturbed by the permanent dipole moment at $t = 0$. 
The light--matter coupling strength $\varepsilon_{k,\lambda}$ between the quantum nuclear dipole moment and the cavity mode $\{k, \lambda\}$ is
\begin{equation}
    \varepsilon_{k,\lambda} = \frac{\omega_{k,\lambda}}{\sqrt{\Omega\epsilon_0}}
    \label{coupling definition}
\end{equation}
Here, $\Omega$ is the effective volume of the cavity, and $\epsilon_0$ is the permittivity of free space. 

The quantum nuclear time-dependent Kohn-Sham matrix $\textbf{F}^{\rm{NEO}}_{\rm n}(t)$ is given by
\begin{equation}
    \textbf{F}^{\rm{NEO}}_{\rm n}(t) = \textbf{H}^{\rm n}_{\rm core}+ \textbf{J}^{\rm nn}[\textbf{P}_{\rm n}(t)] + \textbf{V}^{\rm nn}_{\rm xc}[\textbf{P}_{\rm n}(t)] - \textbf{J}^{\rm ne}[\textbf{P}_{\rm e}(t)] -  \textbf{V}_{\rm c}^{\rm ne}[\textbf{P}_{\rm n}(t),\textbf{P}_{\rm e}(t)] 
    \label{rt neo nuc KS}
\end{equation}
where $\pe$ and $\pn$ are the electronic and quantum nuclear density matrices, respectively, at time $t$. 
In Eq. \ref{rt neo nuc KS}, $\textbf{H}^{\rm n}_{\rm core}$ is the core Hamiltonian describing the quantum nuclear kinetic energy and the Coulomb interaction between quantum and classical nuclei; $\textbf{J}^{\rm nn}[\textbf{P}_{\rm n}(t)]$ describes the Coulomb interaction between quantum nuclei; $\textbf{V}^{\rm nn}_{\rm xc}[\textbf{P}_{\rm n}(t)]$ describes the exchange--correlation interaction between quantum nuclei; $\textbf{J}^{\rm ne}[\textbf{P}_{\rm e}(t)]$ describes the Coulomb interaction between quantum nuclei and electrons; and $\textbf{V}_{\rm c}^{\rm ne}[\textbf{P}_{\rm n}(t),\textbf{P}_{\rm e}(t)]$ describes the correlation energy between quantum nuclei and electrons. 

As mentioned above, we are invoking the electronic Born--Oppenheimer variant of RT-NEO in this work. At each time step, we determine the electronic density matrix $\pe$ by quenching the electronic density to the ground state according to the self-consistent field (SCF) procedure\cite{li_electronic_2023}:  
\begin{equation}
    \textbf{P}_{\rm e}(t) = \textrm{SCF}\left[\Re(\textbf{P}_{\rm n}(t)), \;\textbf{R}_{\rm c}\right]
    \label{bo scf equation}
\end{equation}
Here, $\Re(\textbf{P}_{\rm n}(t))$ is the real part of the quantum nuclear density matrix at time $t$, and $\textbf{R}_{\rm c}$ is the vector of coordinates of the fixed classical nuclei.
This electronic quenching eliminates the need for a small time step to explicitly propagate the electron dynamics, thereby providing a marked reduction in computational cost. 
In previous work\cite{li_electronic_2023, welman_light-matter_2025}, we have shown that this approximation yields excellent agreement with the results obtained via explicit propagation of the electron dynamics when considering VSC on an electronic ground state surface.

At each time step, we obtain the reduced density matrices $\pf$ and $\pn$ for the cavity mode and quantum nuclear subsystems, respectively, by taking the partial trace over the other subsystem:
\begin{equation}
    \textbf{P}_{\rm n}(t) = \textrm{Tr}_{\rm F}[\textbf{P}_{\rm Fn}(t)]
    \label{field partial trace}
\end{equation}
\begin{equation}
    \textbf{P}_{\rm F}(t) = \textrm{Tr}_{\rm n}[\textbf{P}_{\rm Fn}(t)]
    \label{matter partial trace}
\end{equation}
The light--matter entanglement can then be measured by computing the von Neumann entropy, defined as
\begin{equation}
    S(t) = -\textrm{Tr}[\textbf{P}_{\rm n}(t)\ln\textbf{P}_{\rm n}(t)] = -\textrm{Tr}[\textbf{P}_{\rm F}(t)\ln\textbf{P}_{\rm F}(t)]
    \label{entropy}
\end{equation}
The second equality is proved in the Supporting Information of Ref. \citenum{welman_light-matter_2025} and has been confirmed numerically for our code. 
In this work, the von Neumann entropy is computed using the reduced density matrix for the cavity mode subsystem. 

The von Neumann entropy $\s$ is a valid measure of light--matter entanglement provided that the joint light--matter system remains a pure state for all times $t \geq 0$.
In this context, when a system or subsystem is in a pure state, the ket $\ket{\psi}$ describing the state of the system is known with 100\% certainty, i.e., the system or subsystem density operator $\hat{\rho}$ can be written as $\hat{\rho} = \ket{\psi}\bra{\psi}$. 
The joint density matrix $\pfn$ describes a pure state for all times $t \geq 0$ because we initialize it as a separable product of pure state cavity mode and quantum nuclear density matrices (see Simulation Details) and then evolve it unitarily according to Eq. \ref{rt neo nuc eom}. 
Although $\pfn$ is initialized as separable, it does not remain separable throughout the dynamics because the cavity mode and quantum nuclear degrees of freedom become entangled with each other. 
When the two subsystems entangle with each other, they each go from being in a pure state to being in a mixed state.
Thus, rather than knowing the subsystem state ket $\ket{\psi}$ with 100\% certainty, as for a pure state, we only know a set of possible kets $\ket{\psi_i}$, each of which has a probability $p_i$ of being the actual state ket of the subsystem.

$\s$ measures how close the two subsystems are to being separable, which therefore is equivalent to measuring the degree of light--matter entanglement. 
$\s = 0$ implies that $\pfn$ is separable and that there is no light--matter entanglement. 
The maximum possible degree of light--matter entanglement is given by $\smax$  = $\ln(N_{\rm min})$, where $N_{\rm min}$ is the smaller of the cavity mode and molecular basis set sizes. 
This theoretical maximum is not guaranteed to be reached in a dynamical simulation. 
Finally, as discussed in the Introduction, light--matter quantum entanglement is not a prerequisite for the hybridization of light and matter degrees of freedom required to form polaritonic states, which can still occur with a classical treatment of the cavity mode.

To obtain the mfq-RT-NEO equations of motion, we assume that there is no quantum entanglement between the cavity mode and the quantum nuclear degrees of freedom. This assumption is equivalent to assuming that the joint light--matter density matrix $\pfn$ is separable at all times $t \geq 0$:
\begin{equation}
    \textbf{P}_{\rm Fn}(t) = \textbf{P}_{\rm F}(t) \otimes \textbf{P}_{\rm n}(t)
    \label{separability ansatz}
\end{equation}
In this case, Eq. \ref{rt neo nuc eom} separates into two equations of motion, one for $\pf$ and the other for $\pn$: 
\begin{equation}
    i\hbar \frac{\partial}{\partial t}\textbf{P}_{\rm F}(t) = \left[\textbf{H}_{\rm F} + \sum_{k,\lambda}\varepsilon_{k,\lambda}\textbf{q}_{k,\lambda}\mu_{{\rm n}, \lambda}(t), \:\textbf{P}_{\rm F}(t)\right]
    \label{mf rt neo field eom}
\end{equation}
\begin{equation}
    i\hbar \frac{\partial}{\partial t}\textbf{P}_{\rm n}(t) = \left[\textbf{F}^{\rm NEO}_{\rm n}(t) + \sum_{k,\lambda}\varepsilon_{k,\lambda}q_{k,\lambda}(t)\boldsymbol{\mu}_{{\rm n}, \lambda}, \:\textbf{P}_{\rm n}(t)\right] 
    \label{mf rt neo nuc eom}
\end{equation}
Here, $q_{k,\lambda}(t) \equiv \textrm{Tr}[\textbf{P}_{\rm F}(t)\textbf{q}_{k,\lambda}]$ denotes the expectation value of $\hat{q}_{k,\lambda}$ at time $t$, and $\mu_{{\rm n}, \lambda}(t) \equiv \textrm{Tr}[\textbf{P}_{\rm n}(t)\boldsymbol{\mu}_{{\rm n}, \lambda}] - \textrm{Tr}[\textbf{P}_{\rm n}(0)\boldsymbol{\mu}_{{\rm n}, \lambda}]$ denotes the expectation value of the time-dependent quantum nuclear dipole moment operator at time $t$ minus the permanent dipole moment.
Eq. \ref{bo scf equation} for the electronic density matrix remains unchanged. 

To quantify the photon statistics for the cavity mode $\{k, \lambda\}$ obtained by the fq- and mfq-RT-NEO methods, we compute the Mandel Q Parameter, $\mandelq$\cite{Mandel:79}, given by 
\begin{equation}
     Q_{k, \lambda}(t) = \frac{{\rm Tr}(\textbf{P}_{\rm F}(t)\textbf{a}^\dagger_{k, \lambda}\textbf{a}^\dagger_{k, \lambda}\textbf{a}_{k, \lambda}\textbf{a}_{k, \lambda}) - {\rm Tr}(\textbf{P}_{\rm F}(t)\textbf{a}^\dagger_{k, \lambda}\textbf{a}_{k, \lambda})^2}{{\rm Tr}(\textbf{P}_{\rm F}(t)\textbf{a}^\dagger_{k, \lambda}\textbf{a}_{k, \lambda})}
    \label{Mandel Q def 1}
\end{equation}
Here, $\textbf{a}_{k, \lambda}$ ($\textbf{a}^\dagger_{k, \lambda}$) is the matrix representation of the operator $\hat{a}_{k, \lambda}$ ($\hat{a}^\dagger_{k, \lambda}$) which annihilates (creates) a single photon in the cavity mode $\{k, \lambda\}$.
If we introduce $\hat{n}_{k, \lambda} = \hat{a}_{k, \lambda}^\dagger\hat{a}_{k, \lambda}$ as the photon number operator for the cavity mode $\{k, \lambda\}$ and $\textbf{n}_{k, \lambda}$ as its matrix representation, we can rewrite Eq. \ref{Mandel Q def 1} as 
\begin{equation}
    \mandelq = \frac{\langle \hat{n}_{k, \lambda}^2 \rangle_t - \langle \hat{n}_{k, \lambda} \rangle_t^2}{\langle \hat{n}_{k, \lambda} \rangle_t} - 1 = \frac{{\rm var}(\hat{n}_{k, \lambda})_t}{\langle \hat{n}_{k, \lambda} \rangle_t} -1 
    \label{Mandel Q def 2}
\end{equation}
Here, $\langle \hat{n}_{k, \lambda} \rangle_t \equiv {\rm Tr}(\pf \textbf{n}_{k, \lambda})$ is the expectation value of the operator $\hat{n}_{k, \lambda}$ and ${\rm var}(\hat{n}_{k, \lambda})_t \equiv \langle \hat{n}_{k, \lambda}^2 \rangle_t - \langle \hat{n}_{k, \lambda} \rangle_t^2$ is the variance of $\hat{n}_{k, \lambda}$ at time $t$. 
From this point forward, we will adopt the notation $\nav \equiv \langle \hat{n}_{k, \lambda} \rangle_t$ and $\varn \equiv {\rm var}(\hat{n}_{k, \lambda})_t$ for these quantities. 
We have also used the bosonic commutator relation $\left[\hat{a}_{k, \lambda}, \hat{a}^\dagger_{k, \lambda}\right] = 1$ to obtain this expression. 

We can see from Eq. \ref{Mandel Q def 2} that $\mandelq$ measures the deviation of the photon number variance (normalized by the photon number expectation value) from unity. 
A state where $\mandelq = 0$, or $\varn = \nav$, is \textit{Poissonian}, a name that stems from the property of a Poisson distribution that its variance equals its mean. 
The canonical example of a state with Poissonian photon statistics is a coherent state, where the probability of the photon number state with $n$ photons, $|c_{n}|^2$, explicitly satisfies a Poisson distribution. 
This situation corresponds to quasi-classical behavior of the field amplitude and intensity. 
While every coherent state is Poissonian, not every Poissonian cavity mode state is a coherent state.
For example, Poissonian statistics can also occur in a mixed state where the populations of the various photon number states satisfy a Poisson distribution \cite{giovannini_stimulated_2019}.

A \textit{super-Poissonian} state is a cavity mode state where $\varn > \nav$ and $\mandelq$ is therefore positive. 
Meanwhile, a \textit{sub-Poissonian} state is a cavity mode state where $\varn< \nav$. 
The simplest example of a sub-Poissonian state is the single-photon Fock state, which has exactly one photon with no variance in the measured photon number. Because $\varn = 0$ for a Fock state with any number of photons, and variance is a non-negative quantity, it follows that the smallest possible value of $\mandelq$ is $-1$, corresponding to a Fock state. In Section 4, we will consider $\s$ and $\mandelq$ obtained from a set of polaritonic RT-NEO simulations with sub-Poissonian, Poissonian, and super-Poissonian initial states for the cavity mode.

\section{Simulation Details}
Our hierarchy of polaritonic RT-NEO methods is implemented in a developer version of QChem \cite{epifanovsky_software_2021}. Excluding only the quantized cavity mode in the mfq-RT-NEO method, all quantum mechanical degrees of freedom (mode, quantum nuclei, and electrons) are propagated with a modified-midpoint unitary transform (MMUT) time-propagation scheme \cite{li_time-dependent_2005} with an additional predictor-corrector scheme \cite{de_santis_pyberthart_2020} used to control numerical error. In the mfq-RT-NEO method, the quantized cavity mode degrees of freedom are propagated with the time-evolution operator $\exp(-i\textbf{H}t/\hbar)$, where $\textbf{H} = \textbf{H}_{\rm F} + \sum_{k,\lambda}\varepsilon_{k,\lambda}\textbf{q}_{k,\lambda}\mu_{{\rm n}, \lambda}(t)$. 
In both methods, the classical nuclei are fixed at specified geometries.   

The molecular system studied in this work is HCN with a fixed C--N bond length of 1.16 \r{A}. The  hydrogen nucleus (i.e., proton) is treated quantum mechanically with the distance between the protonic basis function center and the carbon nucleus fixed to 1.07 \r{A}. The electronic degrees of freedom were treated with the cc-pVDZ \cite{dunning_gaussian_1989} electronic basis set and the B3LYP \cite{lee_development_1988, becke_new_1998, becke_density-functional_1988} electronic exchange--correlation functional. The quantized proton was treated with the even-tempered 8s8p8d \cite{yang_development_2017} protonic basis set with exponents ranging from $2\sqrt{2}$ to 32, and the epc17-2 \cite{yang_development_2017, brorsen_multicomponent_2017} functional was used to describe electron--proton correlation. 
The cavity mode was described with 15 number basis functions, the largest computationally tractable set given our resources.

At $t = 0$, we chose $\textbf{P}_{\rm F}(0) = \ket{\psi_{\rm F}(0)}\bra{\psi_{\rm F}(0)}$, where $\ket{\psi_{\rm F}(0)}$ was taken to be a coherent (Poissonian), squeezed coherent (super-Poissonian), or single-photon Fock (sub-Poissonian) state. The electronic and quantum nuclear density matrices $\textbf{P}_{\rm e}(0)$ and $\textbf{P}_{\rm n}(0)$ were chosen to be the SCF ground state electronic and quantum nuclear density matrices from a NEO-DFT calculation. For the fq-RT-NEO method, we then initialized the joint density matrix $\textbf{P}_{\rm Fn}(0)$ as a separable product of cavity mode and quantum nuclear density matrices: $\textbf{P}_{\rm Fn}(0) = \textbf{P}_{\rm F}(0) \otimes \textbf{P}_{\rm n}(0)$. We propagated the real-time dynamics with a time step of $\Delta t = 0.04$ a.u. and light--matter coupling of $8 \times 10^{-4}$ a.u. for both methods.

\section{Results}

\begin{figure}[H]
\centering
\includegraphics[scale=0.5]{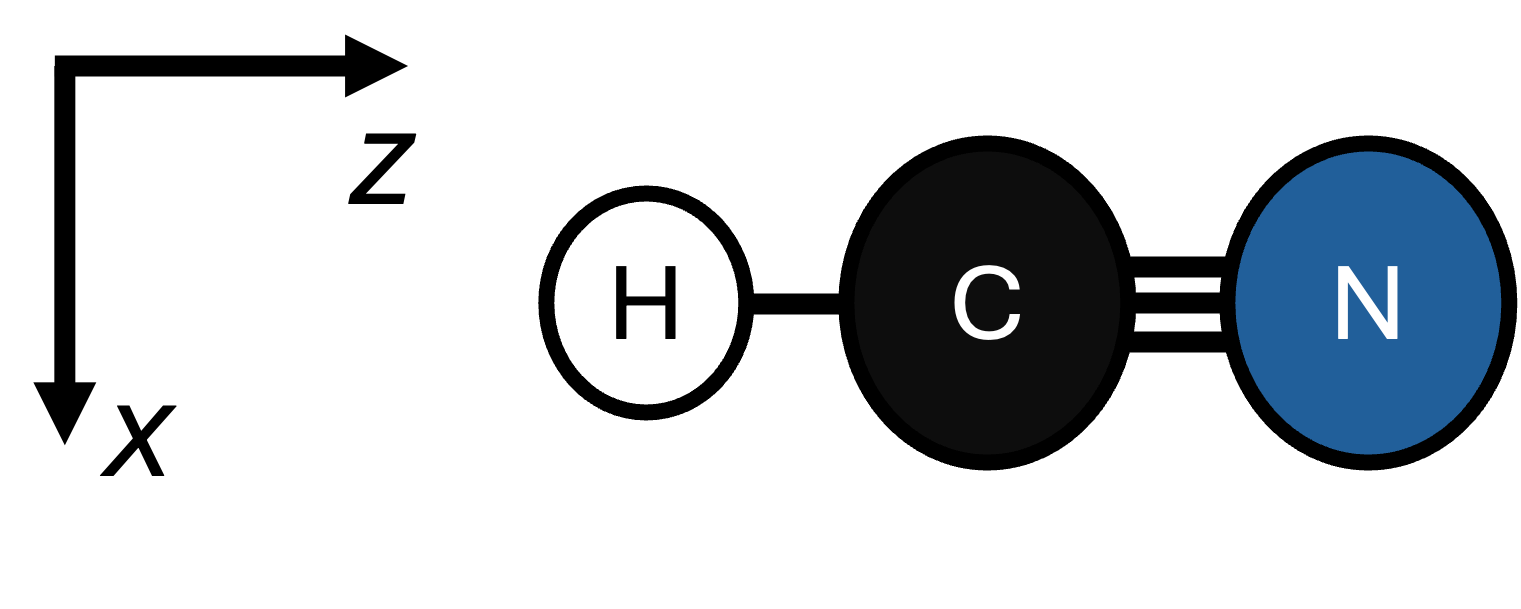}
    \caption{Geometry of the HCN molecule used in the RT-NEO VSC calculations.}
    \label{MoleculeSetup}
\end{figure}

Using the RT-NEO methods described above, we simulated the strong coupling of an optical cavity mode to the HCN molecule shown in Figure \ref{MoleculeSetup}. 
A single $x$-polarized cavity mode was coupled to the $x$-direction bending mode of the quantum proton, which has frequency $\omega_{\rm c}$ = 2803 cm$^{-1}$ at this level of theory, with coupling strength $\varepsilon = 8 \times 10^{-4}$ a.u.

We then propagated the real-time dynamics using the mfq- and fq-RT-NEO methods. The molecular initial state at $t = 0$ is the NEO-DFT ground state, while the cavity mode initial state is one of three states with either Poissonian, super-Poissonian, or sub-Poissonian photon statistics. 
The dynamics obtained with each cavity mode initial state are discussed below. 

\subsection{Coherent State (Poissonian) Initial Condition}

We first consider the photon statistics obtained from the mfq- and fq-RT-NEO methods when the cavity mode initial state $\ket{\psi_{\rm F}(0)}$ is prepared as a coherent state:
\begin{equation}
    \ket{\psi_{\rm F}(0)} = \hat{D}(\alpha)\ket{0}
    = \exp(\alpha\hat{a}^\dagger - \alpha^*\hat{a})\ket{0}
    \equiv \ket{\alpha}
    \label{coherent state definition}
\end{equation}
The density operator associated with this coherent state is the pure state density operator $\hat{\rho}(0) = \ket{\alpha}\bra{\alpha} = \hat{D}(\alpha)\ket{0}\bra{0}\hat{D}^\dagger(\alpha)$. The state $\ket{0}$ is the vacuum state of the single cavity mode (i.e., no photon occupation).
Here, we make use of the displacement operator $\hat{D}(\alpha) = \exp(\alpha\hat{a}^\dagger - \alpha^*\hat{a})$. 
The operator $\hat{a}$ ($\hat{a}^\dagger$) annihilates (creates) a single photon in the cavity mode.
The parameter $\alpha$, which determines the initial displacement of the cavity mode, is in general complex but is assumed to be real throughout this work.
The displacement operator is equivalent to giving the cavity mode coordinate a nonzero expectation value while leaving the cavity mode momentum expectation value zero.
We note that a coherent state is an eigenstate of the annihilation operator $\hat{a}$ with eigenvalue $\alpha$, i.e., $\hat{a}\ket{\alpha} = \alpha\ket{\alpha}$. This property leads directly to the quasi-classical behavior of the coherent state. 

The formula for the coefficients of $\ket{\alpha}$ in the basis of photon number states $\ket{n}$ is given by
\begin{equation}
    \ket{\alpha} = \sum_{n = 0}^{\infty}e^{-|\alpha|^2/2}\frac{\alpha^n}{\sqrt{n!}}\ket{n}
    \label{coherent state coefficients}
\end{equation}
The matrix elements of the associated pure state density matrix $\rho_{mn} = \braket{m|\hat{\rho}|n} = \braket{m|\alpha}\braket{\alpha|n}$ are thus
\begin{equation}
    \rho_{mn} = e^{-|\alpha|^2} \frac{\alpha^{m+n}}{\sqrt{m!n!}}
    \label{coherent state density matrix elements}
\end{equation}
where populations are given by the diagonal elements $m = n$ and coherences by the off-diagonal elements $m \neq n$. The populations $\rho_{nn} =  e^{-|\alpha|^2} |\alpha|^{2n}/n!$ obey a Poisson distribution in the photon number $n$. 
We note here that the populations in the density matrix for a Poissonian mixed state will also obey this distribution, but the coherences will in general differ from those given by Eq. \ref{coherent state density matrix elements} and may even be vanishing. 
To ensure that the expectation value of the coordinate operator for the cavity mode is equal to the initial field amplitude $E_0$, we set $\alpha = \sqrt{\omega_{k, \lambda}/2} E_0$.
In these calculations, we choose an initial field amplitude $E_0 = 0.3$ a.u., which 
corresponds to $\alpha = 0.024$ when the cavity mode frequency $\omega_{k, \lambda}$ is tuned to be resonant with our chosen molecular vibrational transition at 2803 cm$^{-1}$. 
Substituting Eq. \ref{coherent state coefficients} into Eq. \ref{Mandel Q def 2}, the initial value of the Mandel Q Parameter is $Q_{k, x}(0) = 0$.
\subsubsection{Analytical Solution for mfq-RT-NEO}
Before considering the RT-NEO results, we first show analytically that with a coherent state initial condition, the mfq-RT-NEO method predicts that the cavity mode remains in a coherent state for all $t \geq 0$.
Our notation assumes multiple cavity modes but can be easily be reduced to the single cavity mode case used in our simulations. 
Let $\ket{\psi(t)}$ be the state ket of the cavity mode subsystem in the Schr\"odinger picture. 
The von Neumann equation for the cavity mode (Eq. \ref{mf rt neo field eom}) is equivalent to the Schr\"odinger equation
\begin{equation}
    i\hbar \frac{\partial}{\partial t}\ket{\psi(t)} = \left[\hat{H}_{\rm F} + \sum_{k,\lambda}\varepsilon_{k,\lambda}\hat{q}_{k,\lambda}\mu_{{\rm n}, \lambda}(t)\right]\ket{\psi(t)}
    \label{mf rt neo schrodinger}
\end{equation}
The Hamiltonian in Eq. \ref{mf rt neo schrodinger} has two terms. 
We can take the time-independent free-field Hamiltonian $\hat{H}_{\rm F}$ as the zeroth-order Hamiltonian and $\sum_{k,\lambda}\varepsilon_{k,\lambda}\hat{q}_{k,\lambda}\mu_{{\rm n}, \lambda}(t)$ as a time-dependent potential $\hat{V}(t)$. 
Introducing the unitary operator $\hat{U}_{\rm I}(t) \equiv \exp(-i\hat{H}_{\rm F}t/\hbar)$, the state of the cavity mode subsystem in the interaction picture is $\ket{\psi_{\rm I}(t)} \equiv \hat{U}_{\rm I}^\dagger(t)\ket{\psi(t)}$. We can then rewrite Eq. \ref{mf rt neo schrodinger} as a Schr\"odinger equation in the interaction picture as
\begin{equation}
\begin{split}
    i\hbar \frac{\partial}{\partial t}\ket{\psi_{\rm I}(t)} &= \hat{V}_{\rm I}(t)\ket{\psi_{\rm I}(t)} \\
    &= \hat{U}^\dagger _{\rm I}(t)\left[\sum_{k,\lambda}\varepsilon_{k,\lambda}\hat{q}_{k,\lambda}\mu_{{\rm n}, \lambda}(t)\right]\hat{U}_{\rm I}(t)\ket{\psi_{\rm I}(t)} \\
\end{split}
\label{mf rt neo interaction}
\end{equation}
where we have used the standard interaction picture operator definition $\hat{V}_{\rm I}(t) \equiv \hat{U}^\dagger _{\rm I}(t)\hat{V}(t) \hat{U}_{\rm I}(t)$.

Next, we substitute Eq. \ref{free field hamiltonian} and employ the Baker-Hausdorff Lemma 
\begin{equation}
    \exp(i\hat{G}\lambda)\hat{A}\exp(-i\hat{G}\lambda) = \sum_{n = 0}^\infty \left(\frac{i^n \lambda^n}{n!}\right) \left[\hat{G}, \left[\hat{G}, ... \left[\hat{G}, \hat{A}\right]\right]...\right]
\end{equation}
with $\lambda = t/\hbar$, $\hat{G} = \hat{H}_{\rm F}$, and $\hat{A} = \hat{q}_{k, \lambda} \equiv \sqrt{1/2\omega_{k, \lambda}}(\hat{a}_{k, \lambda} + \hat{a}^\dagger_{k, \lambda})$. This procedure leads to 
\begin{equation}
    i\hbar \frac{\partial}{\partial t}\ket{\psi_{\rm I}(t)} =\sum_{k,\lambda}\frac{\varepsilon_{k,\lambda}}{\sqrt{2\omega_{k,\lambda}}}\left(\hat{a}_{k,\lambda}e^{-i\omega_{k, \lambda}t} + \hat{a}^\dagger_{k,\lambda}e^{i\omega_{k, \lambda}t}\right)\mu_{{\rm n}, \lambda}(t) \ket{\psi_{\rm I}(t)} 
\label{mf rt neo interaction 2}
\end{equation}
where we have also invoked the bosonic relation $\left[\hat{a}_{k, \lambda}, \hat{a}^\dagger_{k^\prime, \lambda^\prime}\right] = \delta_{kk^\prime}\delta_{\lambda\lambda^\prime}$. 
Eq. \ref{mf rt neo interaction 2} may be solved iteratively as a Dyson series. The result is 
\begin{equation}
    \ket{\psi_{\rm I}(t)} = \textbf{T}\exp\left(-\frac{i}{\hbar}\sum_{k,\lambda}\int_0^t dt^\prime \frac{\varepsilon_{k,\lambda} \mu_{{\rm n}, \lambda}(t^\prime)}{\sqrt{2\omega_{k,\lambda}}}\left[\hat{a}_{k,\lambda}e^{-i\omega_{k, \lambda}t^\prime} + \hat{a}^\dagger_{k,\lambda}e^{i\omega_{k, \lambda}t^\prime}\right]\right)\ket{\psi(0)}
    \label{dyson series}
\end{equation}
where we have used the fact that $\ket{\psi_{\rm I}(0)} = \ket{\psi(0)}$ (i.e., the state at $t = 0$ is picture-independent) and $\textbf{T}$ denotes time ordering. 
Time ordering is necessary here because for a given mode $\{k, \lambda\}$, the term within the square  brackets does not commute with itself at different times. 
If we now define a time-dependent displacement parameter $\alpha_{k,\lambda} (t)$ as
\begin{equation}
    \alpha_{k,\lambda} (t) = -\frac{i}{\hbar}\int_0^t dt^\prime \frac{\varepsilon_{k,\lambda} \mu_{{\rm n}, \lambda}(t^\prime)}{\sqrt{2\omega_{k,\lambda}}}e^{i\omega_{k, \lambda}t^\prime}
    \label{alpha definition}
\end{equation}
then by once again invoking $\left[\hat{a}_{k, \lambda}, \hat{a}^\dagger_{k^\prime, \lambda^\prime}\right] = \delta_{kk^\prime}\delta_{\lambda\lambda^\prime}$ we can rewrite Eq. \ref{dyson series} as 
\begin{equation}
\begin{split}
    \ket{\psi_{\rm I}(t)} &= \textbf{T}\exp\left[\sum_{k, \lambda}\left(\alpha_{k,\lambda}(t)\hat{a}^\dagger_{k,\lambda} - \alpha^*_{k,\lambda}(t)\hat{a}_{k,\lambda}\right)\right]\ket{\psi(0)} \\
    &= \textbf{T}\prod_{k, \lambda}\hat{D}_{k, \lambda}(\alpha_{k, \lambda}(t))\ket{\psi(0)}
    \label{time order displacement no initial condition}
\end{split}
\end{equation}
Here, the displacement operator $\hat{D}_{k, \lambda}(\alpha_{k, \lambda}(t))$ is defined in parallel with Eq. \ref{coherent state definition} as
\begin{equation}
    \hat{D}_{k, \lambda}(\alpha_{k, \lambda}(t)) = \exp\left(\alpha_{k,\lambda}(t)\hat{a}_{k,\lambda}^\dagger - \alpha^*_{k,\lambda}(t)\hat{a}_{k,\lambda}\right).
    \label{displacement operator}
\end{equation} 
Finally, because all cavity modes are initialized as coherent states with initial displacements $\alpha_{k, \lambda}(0)$, all of which are assumed to be real, we can write Eq. \ref{time order displacement no initial condition} as
\begin{equation}
    \ket{\psi_{\rm I}(t)} = \textbf{T}\prod_{k, \lambda}\hat{D}_{k, \lambda}(\alpha_{k, \lambda}(t))\hat{D}_{k, \lambda}(\alpha_{k, \lambda}(0))\ket{\textbf{0}}
    \label{final form}
\end{equation}
Here, $\ket{\textbf{0}}$ is the state where all cavity modes are in the vacuum state (i.e., occupied by zero photons). 

Eq. \ref{final form} shows that at any time $t \geq 0$, the total cavity mode subsystem state in the interaction picture can be written as a product of time-ordered displacement operators, one for each cavity mode, acting on the vacuum state $\ket{\textbf{0}}$ of the complete set of quantized cavity modes. 
Since the product of displacement operators is itself a displacement operator, and a coherent state is produced by acting a displacement operator on the vacuum state according to Eq. \ref{coherent state definition}, this equation implies that all cavity modes are in coherent states at all times $t \geq 0$ in the interaction picture. 
Furthermore, this result is picture-independent: from the definition $\ket{\psi_{\rm I}(t)} \equiv \hat{U}_{\rm I}^\dagger(t)\ket{\psi(t)}$, we can see that the cavity mode state in the Schr\"odinger picture is the state in the  interaction picture evolved forward in time under the influence of the free field Hamiltonian given in Eq. \ref{free field hamiltonian} by an amount $t$. 
Any cavity mode initially in a coherent state will remain in a coherent state under forward or backward time evolution with the Hamiltonian given in Eq. \ref{free field hamiltonian}. Thus, any cavity mode that is in a coherent state in the interaction picture will also be in a coherent state in the Schr\"odinger picture. 
Several results immediately follow from this finding. 
For example, this analysis shows that $\mandelq = 0$ (i.e., ${\rm var}\left(\hat{n}_{k, \lambda}\right)_t = \langle \hat{n}_{k, \lambda} \rangle_t$) for all $\{k, \lambda\}$ and $t \geq 0$ when calculated with the mfq-RT-NEO method. 
We confirm this result and associated properties numerically for a single cavity mode below. 

\subsubsection{Numerical Results for mfq-RT-NEO and fq-RT-NEO}
The fq-RT-NEO results should capture deviations from the mean-field result derived above due to light--matter entanglement. 
Before considering the photon statistics, we first identify the key timescales associated with our coherent state initial condition. 

\begin{figure}[H]
\centering
\includegraphics[scale=0.35]{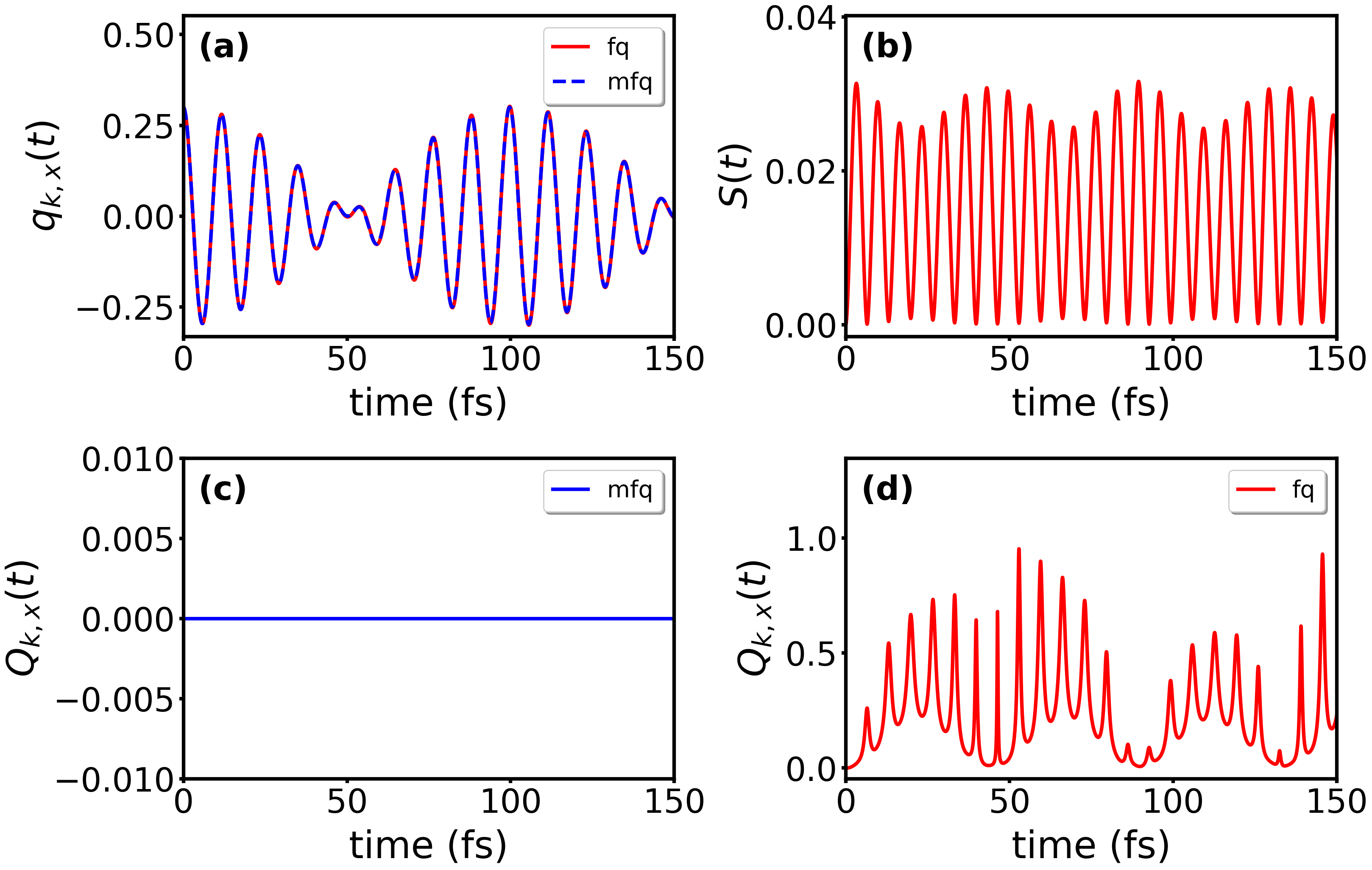}
    \caption{Time-dependent quantities obtained from mfq- and fq-RT-NEO dynamics with a coherent state cavity mode initial condition. (a) Cavity mode coordinate $q_{k, x}(t)$. (b) fq-RT-NEO von Neumann entropy $\s$. (c) mfq-RT-NEO Mandel Q Parameter $\mandelqx$. (d) fq-RT-NEO Mandel Q Parameter $\mandelqx$.}
    \label{coherent_1}
\end{figure}

Figure \ref{coherent_1}a shows the cavity mode coordinate $q_{k, x}(t) \equiv {\rm Tr}\left[\textbf{P}_{\rm F}(t)\textbf{q}_{k, x}\right]$ calculated with the mfq- and fq-RT-NEO methods. 
As expected from our previous work,\cite{welman_light-matter_2025} the two time-domain signals are in excellent agreement.
We identify two timescales from the dynamics of the cavity mode coordinate. The timescale $\tc \simeq $ 12 fs is the period of the fast oscillations of $q_{k, x}(t)$. 
This timescale corresponds to the cavity mode frequency $\omega_{\rm c}$ = 2803 cm$^{-1}$ according to $\tc = 2\pi/\omega_{\rm c}$. 
The fast oscillations are modulated by slow oscillations on the timescale $\tr \simeq$ 103 fs, which corresponds to the period of the Rabi oscillations. 
We note that the Rabi frequency $\Omega_{\rm R}$ discussed in our previous work is related to this quantity by $\Omega_{\rm R} = 2\pi/\tr$.
Figure \ref{coherent_1}b shows the dynamics of the von Neumann entropy $\s$ calculated with the fq-RT-NEO method. 
Fast oscillations are observed to occur with a period $\tc/2 \simeq$ 6 fs, modulated by a slower oscillation with period $\te \simeq 45$ fs.
This slower period is roughly half of $\tr$, a result that can be anticipated from the Jaynes-Cummings model, as discussed in our previous work \cite{welman_light-matter_2025}. 
The entanglement Rabi frequency $\Omega_{\rm E}$ discussed in our previous work is related to this quantity by $\Omega_{\rm E} = 2\pi/\te$.

After identifying the key timescales in $q_{k, x}(t)$ and $\s$, we turn our attention to the photon statistics. 
Figure \ref{coherent_1}c shows the Mandel Q Parameter $\mandelqx$ calculated with the mfq-RT-NEO method. 
As expected from the analytical result shown above, the mfq-RT-NEO method gives a constant $\mandelqx = 0$. 
In contrast, the fq-RT-NEO result in Figure \ref{coherent_1}d shows significant super-Poissonian fluctuations with a maximum value of $\sim 0.95$. 
The timescales of both the fast and slow oscillations of $\mandelqx$ match those of $\s$, with fast oscillations of period $\tc/2$ modulated by slow oscillations of period $\te$.
From this point forward, we will refer to the maxima (minima) of the fast oscillations as the ``fast maxima (minima)" and the maxima (minima) of the slow oscillations as the ``slow maxima (minima)".
The fast oscillations of $\s$ and $\mandelqx$ are 180 degrees out of phase, indicating that a fast maximum of $\s$ corresponds to a fast minimum of $\mandelqx$ and vice  versa. 
This observation suggests that the cavity mode is oscillating with fast oscillation period $\tc/2$ between two different types of states. 

The fast minima of $\s$ have very low values only slightly greater than 0.0.
We will therefore refer to these states as ``weakly mixed" states. 
We note that the magnitude of $\s$ in these weakly mixed states displays small slow oscillations with period $\te$. 
The fast minima of $\s$ correspond to fast maxima of the Mandel Q Parameter that have $\mandelqx > 0$.
This observation indicates that the weakly mixed cavity mode states all have super-Poissonian photon statistics.

In contrast, the fast maxima of $\s$ have values that are consistently positive and also exhibit slow oscillations with period $\te$. 
We will refer to these states as ``strongly mixed" states, as their von Neumann entropies are generally at least an order of magnitude larger than those of the weakly mixed states discussed above.
In contrast to the weakly mixed states, the regime of photon statistics for the strongly mixed states oscillates as a function of time. 
By comparing the fast oscillations of $\s$ and $\mandelqx$, we can identify two different subclasses of strongly mixed states with different photon statistics. 
The strongly mixed states correspond to fast minima of the Mandel Q Parameter $\mandelqx$. 
Most of these states have $\mandelqx > 0$, indicating that they have super-Poissonian photon statistics.
On a period of $\te$, however, the value of the Mandel Q Parameter returns to a slow minimum of $\mandelqx \sim 0$. 
In this case, the strongly mixed states have Poissonian photon statistics. 
Therefore, these states must be Poissonian mixed states where the density matrix populations obey a Poisson distribution. 
The observation that these states also correspond to a slow maximum of $\s$ indicates that the Poissonian mixed states are also the most mixed of the strongly mixed states. 

\begin{figure}[H]
\centering
\includegraphics[scale=0.35]{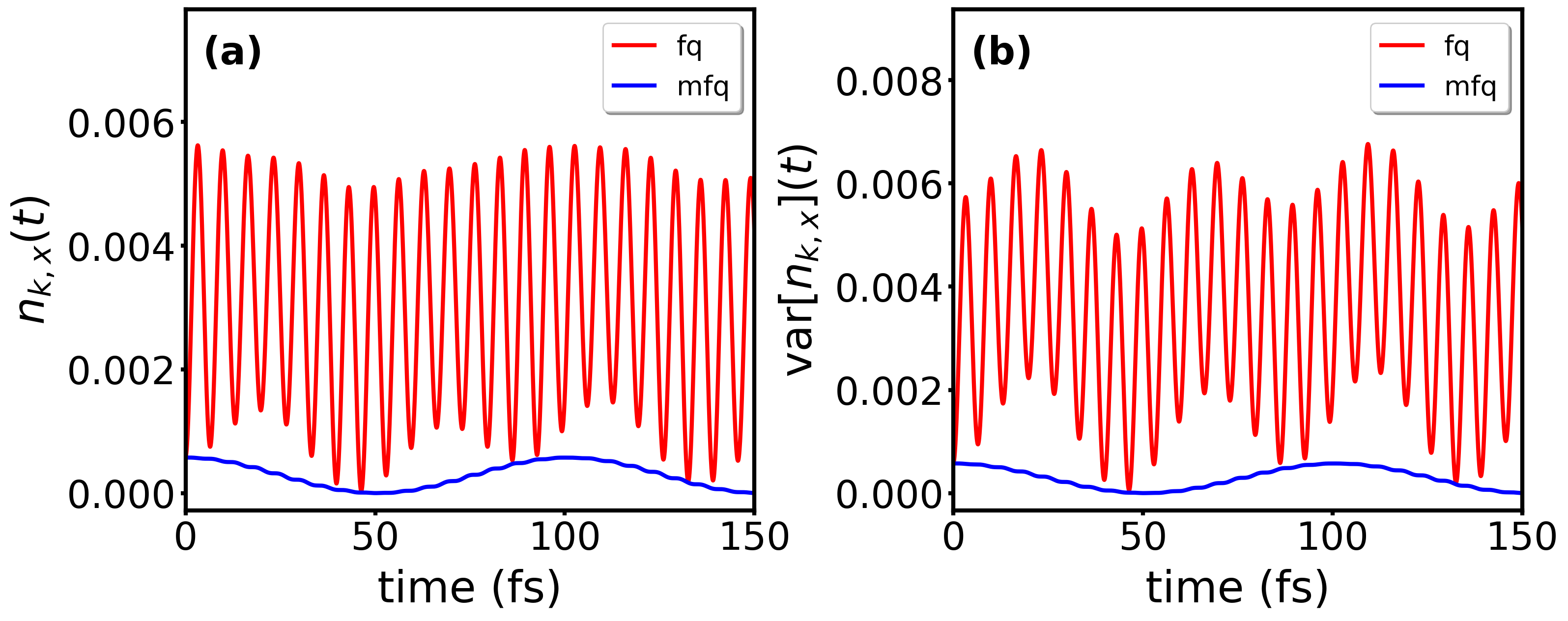}
    \caption{Photon number expectation value and variances obtained from mfq-RT-NEO and fq-RT-NEO dynamics with a coherent state cavity mode initial condition. (a) Photon number expectation value $\nav$. (b) Photon number variance $\varn$.}
    \label{coherent_2}
\end{figure}

To further our analysis of $\mandelqx$ shown in Figures \ref{coherent_1}c and d, we can decompose this quantity into its constituent terms, namely the photon number expectation value $\nav$ and the photon number variance $\varn$ (see Eq. \ref{Mandel Q def 2}). 
Figure \ref{coherent_2}a shows the dynamics of $\nav$ calculated with the mfq- and fq-RT-NEO methods. 
In the mfq-RT-NEO method, weak fast oscillations of period $\tc/2$ are modulated by slow sinusoidal oscillations with period $\tr$. 
The dynamics of $\nav$ calculated with the fq-RT-NEO method, on the other hand, show significantly larger fast oscillations than those calculated with the mfq-RT-NEO method.
Moreover, the fast oscillations with period $\tc/2$ are modulated by slow oscillations on two distinct timescales. 
The values of the fast maxima oscillate on a timescale $\tr$, while the values of the fast minima oscillate on a timescale $\te$.
Interestingly, $\nav$ increases at the beginning of the trajectory, implying that the interaction energy between the cavity mode and the molecule decreases given that the molecule is initialized in its ground state and the total energy is conserved. 

Figure \ref{coherent_2}b shows the dynamics of $\varn$ calculated with the mfq- and fq-RT-NEO methods. 
As expected from the analytical proof in Section 4.1.1, the mfq-RT-NEO result for $\varn$ is in excellent agreement with that for the corresponding result for $\nav$. 
The fq-RT-NEO result for $\varn$ also shows parallels with the corresponding result for $\nav$. 
In this case, however, fast oscillations with period $\tc/2$ are modulated by slow oscillations on a single timescale $\te$. 
The fast oscillations of $\nav$ and $\varn$ in the fq-RT-NEO calculations are in phase with each other and with those of $\s$ but are out of phase with those of $\mandelqx$. 
This observation suggests that the weakly mixed states have lower photon number expectation values and variances than the strongly mixed states. 

To summarize this analysis of $\s$, $\mandelqx$, and the photon number statistics, there are two different classes of states appearing in the dynamics with a coherent state initial condition. 
The first class is the weakly mixed states, all of which have super-Poissonian photon statistics. 
The second class is the strongly mixed states, which can be separated into two subclasses, each with different photon statistics. 
Most of the strongly mixed states have super-Poissonian statistics, but Poissonian statistics recur on a timescale $\te$.
The Poissonian strongly mixed states have the largest von Neumann entropy of any of the strongly mixed states. 
Finally, the weakly mixed states have lower photon number expectation values and variances than the strongly mixed states.

In addition to photon statistics and von Neumann entropy, we can also look at the dynamics of observables associated with the cavity mode quadrature operators $\xl$ and $\yl$, which are defined by
\begin{equation}
\begin{split}
    \hat{a}_{k, \lambda} = \xl + i\yl \\
    \hat{a}^\dagger_{k, \lambda} = \xl - i\yl
    \label{axy def}
\end{split}
\end{equation}
For the cavity mode coordinate operator $\hat{q}_{k, \lambda} = \sqrt{1/2\omega_{k, \lambda}}(\hat{a}_{k, \lambda} + \hat{a}^\dagger_{k, \lambda})$ and cavity momentum operator $\hat{p}_{k, \lambda} = -i\sqrt{\omega_{k, \lambda}/2}(\hat{a}_{k, \lambda} - \hat{a}^\dagger_{k, \lambda})$, we can write $\xl$ and $\yl$ as 
\begin{equation}
\begin{split}
    \xl = \sqrt{\frac{\omega_{k, \lambda}}{2}}\hat{q}_{k, \lambda} \\
    \yl = \sqrt{\frac{1}{2\omega_{k, \lambda}}}\hat{p}_{k, \lambda} 
\end{split}
\label{qpxy}
\end{equation}
In other words, $\xl$ and $\yl$ are dimensionless versions of the cavity mode coordinate and momentum operators, respectively.
For a coherent state, the variances of the quadrature operators can be shown to be equal. If $\textbf{P}_{\rm F}(t)$ is the density matrix for a coherent state, then we can show that 
\begin{equation}
\begin{split}
        \varxl &= {\rm Tr}\left[\textbf{P}_{\rm F}(t)\textbf{X}_{k, \lambda}^2\right] - {\rm Tr}\left[\textbf{P}_{\rm F}(t)\textbf{X}_{k, \lambda}\right]^2 = \frac{1}{4}\\
        \varyl &= {\rm Tr}\left[\textbf{P}_{\rm F}(t)\textbf{Y}_{k, \lambda}^2\right] - {\rm Tr}\left[\textbf{P}_{\rm F}(t)\textbf{Y}_{k, \lambda}\right]^2 = \frac{1}{4}\\
\end{split}
\label{coherent variance}
\end{equation}
where $\textbf{X}_{k, \lambda}$ and $\textbf{Y}_{k, \lambda}$ are the matrix representations of $\xl$ and $\yl$. The quadrature operator variances are thus equal for a coherent state. 
Equality does not hold for a general cavity mode state, which typically exhibits unequal quadrature variances. 
In this situation, the light is considered to be ``quadrature squeezed" in the direction with smaller variance.  
This phenomenology is a sign that the quantum state of the cavity mode has deviated from a coherent state and will be considered extensively in our subsequent analysis. 
Furthermore, it immediately follows that the product of the quadrature variances obeys the uncertainty relation
\begin{equation}
    \varxl \varyl \geq \frac{1}{16}
    \label{uncertainty principle}
\end{equation}
which follows from substituting Eq. \ref{qpxy} into the commutator relation $\left[\hat{q}_{k, \lambda}, \hat{p}_{k^\prime, \lambda^\prime} \right] = i\delta_{kk^\prime}\delta_{\lambda\lambda^\prime}$ and invoking the general uncertainty relation. 
Combining Eqs. \ref{coherent variance} and \ref{uncertainty principle}, we can see that a coherent state is a \textit{minimum uncertainty state} in the cavity mode quadratures by virtue of the fact that the product of its variances satisfies the equality in the uncertainty relation.

\begin{figure}[H]
\centering
\includegraphics[scale=0.35]{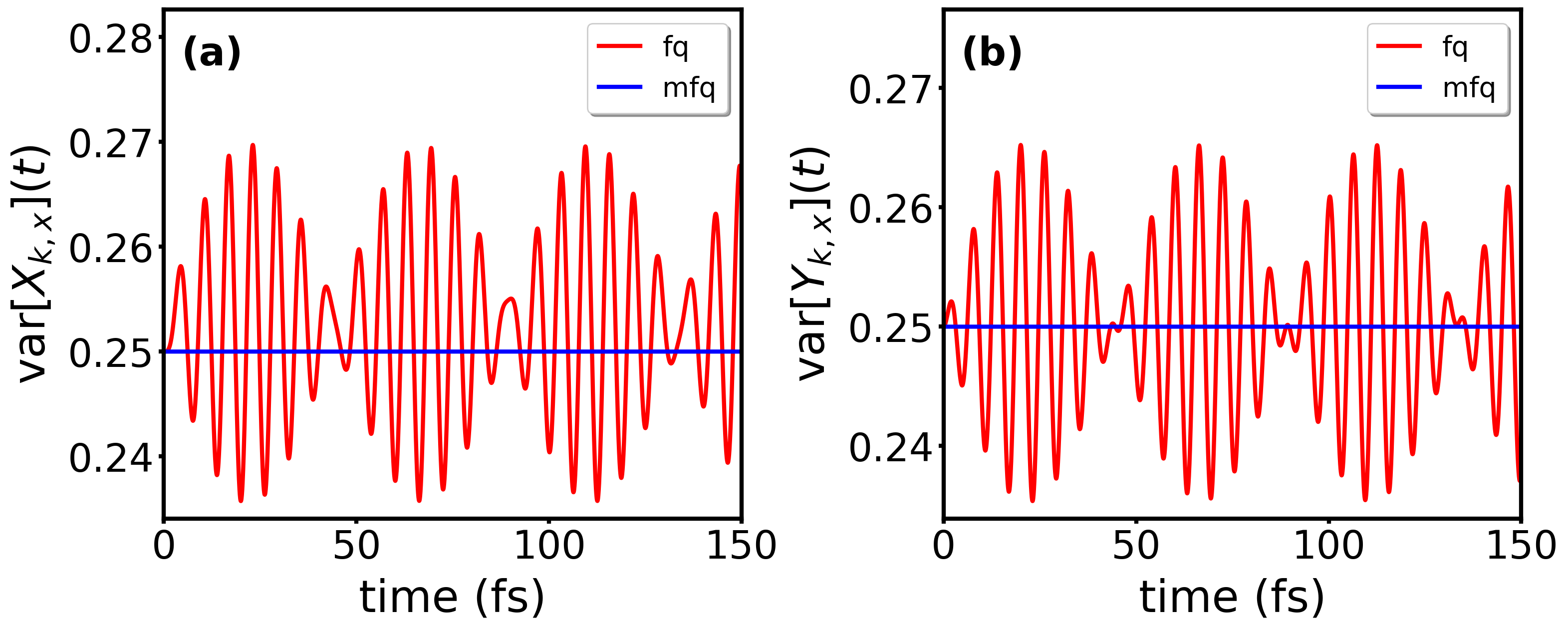}
    \caption{Cavity mode quadrature variances obtained from mfq-RT-NEO and fq-RT-NEO dynamics with a coherent state cavity mode initial condition. (a) \textit{X}-quadrature variance $\varxx$. (b) \textit{Y}-quadrature variance $\varyx$. }
    \label{coherent_3}
\end{figure}

We now consider the quadrature variances, $\varxx$ and $\varyx$, from our simulations. 
These variances are shown in Figures \ref{coherent_3}a and b, respectively.
As expected from the analytical result in Section 4.1.1. combined with Eq. \ref{coherent variance}, the mfq-RT-NEO method predicts both quadrature variances to be a constant 1/4. 
In contrast, the quadrature variances calculated with the fq-RT-NEO method show Rabi oscillations with period $\te$. 
These slow oscillations, which are in phase between the two variances, modulate fast oscillations that have period $\tc/2$ and are approximately 180 degrees out of phase between the two variances. 
An increase in one quadrature variance therefore corresponds to a decrease in the other by the same magnitude.
The fast oscillations of $\varxx$ are in phase with those of $\s$.
This observation indicates that the fast maxima of $\s$ correspond to increased values of $\varxx$ and decreased values of $\varyx$ relative to the coherent state value of 1/4, and vice versa. 
These results demonstrate that the weakly mixed states are \textit{X}-quadrature squeezed, while the strongly mixed states are \textit{Y}-quadrature squeezed.
This squeezing occurs as a result of light--matter entanglement between the cavity mode and the molecular vibrational mode. 

We summarize the full picture  of the dynamics observed in this section as follows. 
As demonstrated analytically, the cavity mode in the mfq-RT-NEO method remains in a coherent state throughout the real-time dynamics, with the photon number expectation value and photon number variance oscillating on the same fast and slow timescales as the cavity mode coordinate. 
By including light--matter entanglement, the fq-RT-NEO results demonstrate significant deviations from the mean-field dynamics. 
This result is particularly interesting, as it shows that even a quasi-classical coherent state can still show noticeable quantum fluctuations if light--matter entanglement is taken into account. 
The cavity mode oscillates between weakly mixed states and strongly mixed states with a period corresponding to half the period of the cavity mode oscillations. 
The weakly mixed states always have super-Poissonian photon statistics.
The strongly mixed states are usually super-Poissonian as well but become Poissonian with a period of $\te$. 
The weakly mixed states are characterized by lower photon number and photon number variance than the strongly mixed states. 
Finally, the weakly mixed states are \textit{X}-quadrature squeezed, while the strongly mixed states are \textit{Y}-quadrature squeezed. 
As demonstrated by comparison to the mfq-RT-NEO results, this quadrature squeezing is the direct result of light--matter entanglement. 

Our result here is consistent with previous work related to the formation of squeezed states in the Jaynes-Cummings model\cite{meystre_squeezed_1982, carmichael_photon_1985}, and the squeezing of light by an emitter in an optical cavity has been demonstrated experimentally in Ref. \citenum{ourjoumtsev_observation_2011}.
It is worth noting, however, that the emitter used in this experiment was a rubidium atom, and the cavity mode was coupled to a transition between two of that atom's electronic states.
Here, we have used a first-principles method incorporating nuclear quantum effects to demonstrate that the squeezing of quantum light can also be achieved by coupling an optical cavity mode to a transition between quantized vibrational states on a ground-state electronic surface in a molecular system. 
Thus, this work provides an interesting generalization of the experimental result.

\subsection{Squeezed Coherent State (Super-Poissonian) Initial Condition}

After finding evidence for light squeezing by the molecular subsystem when the cavity mode is prepared in a coherent state, we now ask what dynamics occur if the initial cavity mode state itself is squeezed. 
We consider the photon statistics obtained from the mfq- and fq-RT-NEO methods when the cavity mode initial state $\ket{\psi_{\rm F}(0)}$ is prepared as a squeezed coherent state:
\begin{equation}
    \ket{\psi_{\rm F}(0)} = \hat{D}(\alpha)\hat{S}(\xi)\ket{0}
    \label{squeezed coherent state def}
\end{equation}
The density operator associated with this squeezed coherent state is the pure state density operator $\hat{\rho}(0) =  \hat{D}(\alpha)\hat{S}(\xi)\ket{0}\bra{0}\hat{S}^\dagger(\xi)\hat{D}^\dagger(\alpha)$. 
The squeeze operator $\hat{S}(\xi)$ is defined as
\begin{equation}
    \hat{S}(\xi) = \exp\left(\frac{1}{2}\xi^*\hat{a}^2 - \frac{1}{2}\xi(\hat{a}^\dagger)^2\right)
    \label{squeeze operator definition}
\end{equation}
where $\xi$ is a complex scalar that will be taken to be real throughout this work. 
The effect of the squeeze operator can be understood as follows. 
As discussed in the previous section, a coherent state is a minimum uncertainty state in the cavity mode quadratures. 
Both quadrature uncertainties are equal, and their product is $1/16$, the minimum bound required by the uncertainty principle.
The action of the squeeze operator $\hat{S}(\xi)$ before the displacement operator $\hat{D}(\alpha)$ eliminates the equality between the two quadrature uncertainties without altering the expectation value of either quadrature that would be obtained by the action of $\hat{D}(\alpha)$ alone \cite{loudon_quantum_2010}. 
In particular, an increasingly large real-valued $\xi$ will decrease the uncertainty in the cavity mode coordinate while increasing the uncertainty in the cavity mode momentum. 

The difference between the photon number variance $\varnl$ and the photon number expectation value $\navl$ in a squeezed coherent state of the cavity mode $\{k, \lambda\}$ can be written in terms of $\alpha$ and $\xi$ according to the formula\cite{loudon_quantum_2010}
\begin{equation}
    \varnl - \navl = \alpha^2(e^{-2\xi} - 1) + 2\sinh^4{\xi} + \sinh^2{\xi}
    \label{squeeze var diff formula}
\end{equation}
which is valid for real $\alpha$ and $\xi$. The squeezed coherent state will be sub-Poissonian for negative values of this difference, Poissonian for zero values, and super-Poissonian for positive values. 
The parameters for the initial squeezed coherent state in the calculations performed here were chosen to be $\alpha = 0.024$, matching the value used for the coherent state calculations in the previous section, and $\xi = 0.05$. 
With these values, $\varn - \nav > 0$ according to Eq. \ref{squeeze var diff formula}, and the initial squeezed coherent state is super-Poissonian with $Q_{k, x}(0) = 0.799$. 

\begin{figure}[H]
\centering
\includegraphics[scale=0.35]{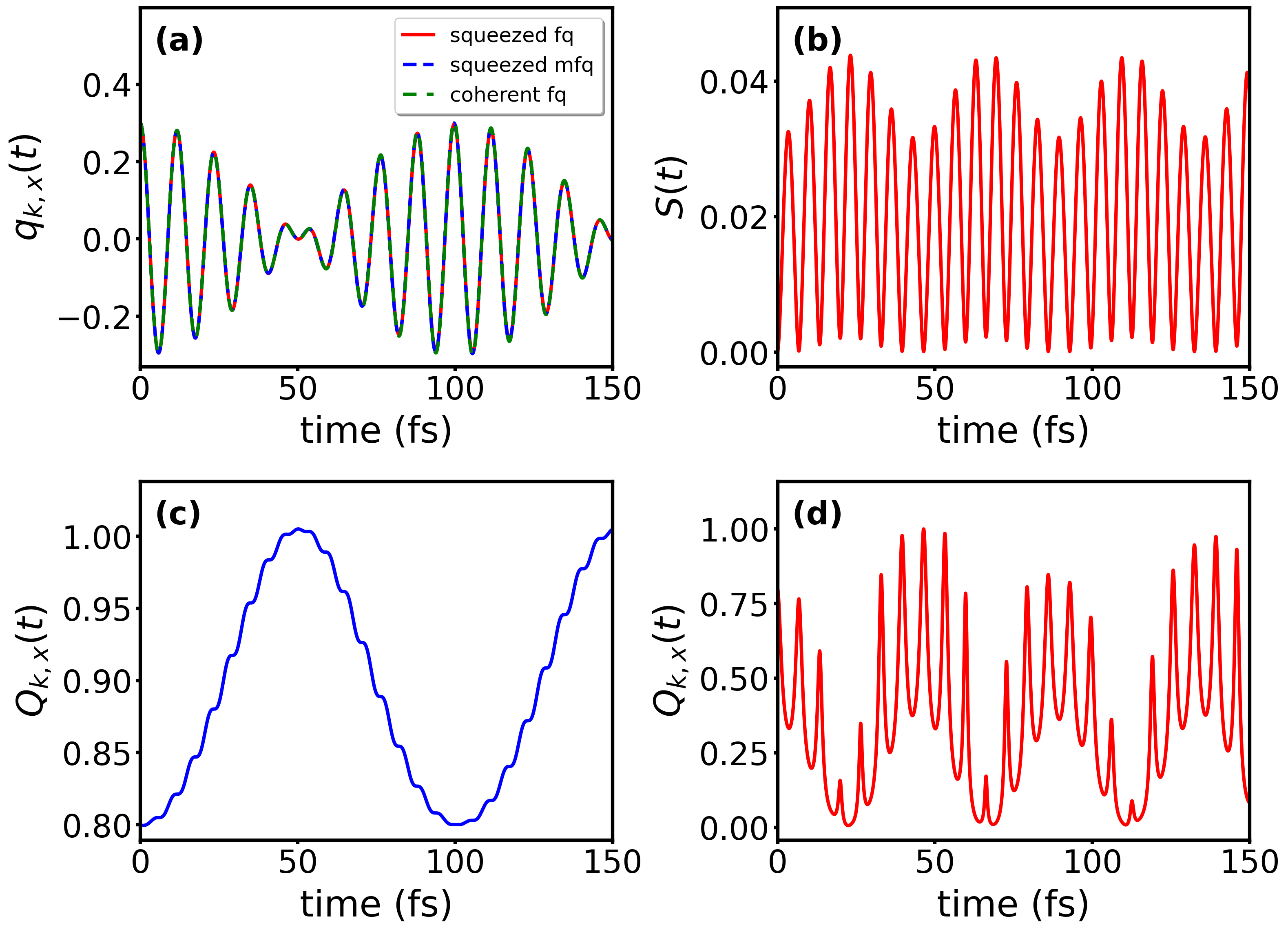}
    \caption{Time-dependent quantities obtained from mfq- and fq-RT-NEO dynamics with a squeezed coherent state cavity mode initial condition. (a) Cavity mode coordinate $q_{k, x}(t)$, with the analogous fq-RT-NEO result for a coherent state initial condition in green. (b) fq-RT-NEO von Neumann entropy $\s$. (c) mfq-RT-NEO Mandel Q Parameter $\mandelqx$. (d) fq-RT-NEO Mandel Q Parameter $\mandelqx$.}
    \label{squeeze_1}
\end{figure}

We first examine the relevant timescales of the various quantities. Figure \ref{squeeze_1}a shows the dynamics of the cavity mode coordinate $q_{k, x}(t)$ obtained with the mfq- and fq-RT-NEO methods. 
The two time domain signals are in excellent agreement with the dynamics of $q_{k, x}(t)$ obtained using the fq-RT-NEO method with a coherent state initial condition that has the same displacement parameter $\alpha$ as the squeezed state.
This satisfying result is consistent with the property that the squeeze operator $\hat{S}(\xi)$ does not alter either of the quadrature expectation values from those that result solely by the action of the displacement operator $\hat{D}(\alpha)$. This property signifies that the value of $q_{k, x}(0)$ is unchanged by the squeezing. 
The equivalence of the time signals indicates that the values of $\tr$ and $\tc$ are unchanged from the values considered in the previous section. 
Figure \ref{squeeze_1}b shows the dynamics of the von Neumann entropy $\s$ obtained with the fq-RT-NEO method. 
It is interesting to note that the maximum value of $\s$ shown here is $\sim 40\%$ larger than the maximum value obtained in Figure \ref{coherent_1}b for the coherent state initial condition.
The maxima of the slow oscillations are also phase-shifted by 180 degrees from the coherent state case.
The implicated timescales, however, are unchanged from the previous results.
Once again, we can identify fast oscillations on a timescale  $\tc/2$ being modulated by slow oscillations on a timescale $\te$. 
The value of $\te$ here is the same as that identified previously for the coherent state initial condition.

We now turn our attention to the photon statistics. 
Figure \ref{squeeze_1}c shows $\mandelqx$ obtained with the mfq-RT-NEO method.
Weak fast oscillations with period $\tc/2$ are modulated by slow oscillations with amplitude $\sim 0.1$ and period $\tr$. 
These slow oscillations result in an increase of $\mandelqx$ from its initial value of 0.799, returning to that value with period $\tr$. 
We note that these dynamics are qualitatively different from the the mfq-RT-NEO result with a coherent state initial condition shown previously in Figure \ref{coherent_1}c. In that case, $\mandelqx$ is a constant for all $t$. 

The mfq-RT-NEO dynamics of $\mandelqx$ with a squeezed coherent state initial condition are also qualitatively different from the corresponding fq-RT-NEO result shown in Figure \ref{squeeze_1}d. 
In the fq-RT-NEO result, $\mandelqx$ is less than the initial value of 0.799 for most times.
Fast oscillations on a timescale $\tc/2$ are modulated by slow oscillations on a timescale $\te$. 
The amplitude of the slow oscillations is $\sim 5$ times the amplitude of the slow oscillations in the mfq-RT-NEO result.
For most of the dynamics, $\mandelqx > 0$, meaning that the cavity mode is in a super-Poissonian state. 
The cavity mode state, however, returns to $\mandelqx = 0$, or Poissonian statistics, on the timescale $\te$.

The timescale behavior of $\mandelqx$ in Figure \ref{squeeze_1}d closely parallels that of $\mandelqx$ for the fq-RT-NEO dynamics with a coherent state initial condition shown in Figure \ref{coherent_1}d. 
By comparison of the locations of the fast and slow minima and maxima of $\s$ and $\mandelqx$ shown in Figures \ref{squeeze_1}b and \ref{squeeze_1}d, we can once again identify two types of states between which the cavity mode oscillates. 
In parallel to our results for a coherent state initial condition, we can identify fast minima of $\s$ as weakly mixed states. 
These weakly mixed states correspond to fast maxima of the Mandel Q Parameter with $\mandelqx > 0$, indicating that they are all super-Poissonian states. 
Similar to the weakly mixed states that were identified with a coherent state initial condition, the values of $\s$ for the weakly mixed states oscillate with period $\te$.  
Furthermore, the fast maxima of $\s$ can be identified as strongly mixed states. 
The strongly mixed states are usually super-Poissonian with $\mandelqx > 0$, but strongly mixed states with Poissonian statistics also occur with period $\te$. 
As for the coherent state initial condition, the Poissonian mixed states are the most mixed of the strongly mixed states. 

\begin{figure}[H]
\centering
\includegraphics[scale=0.35]{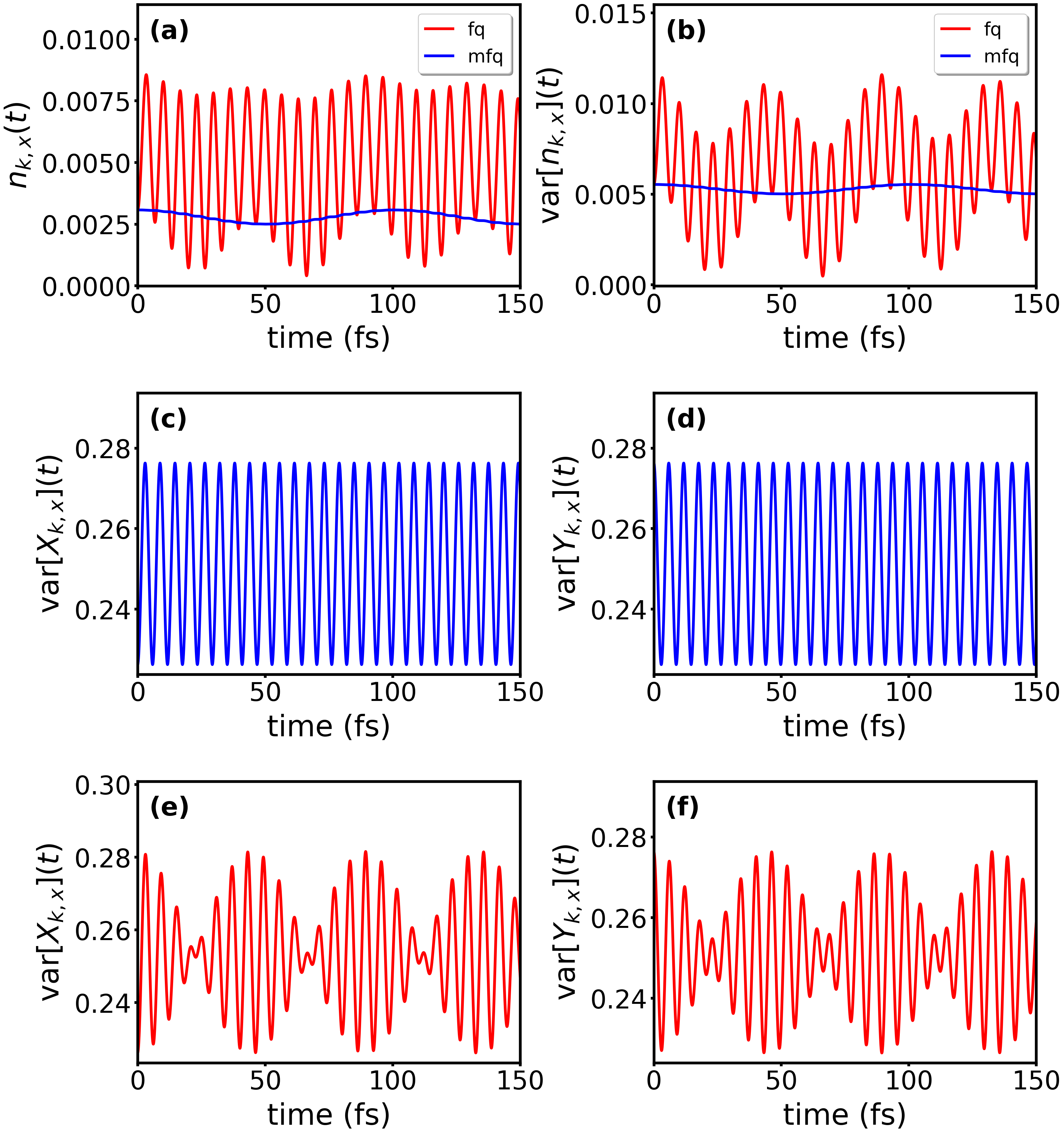}
    \caption{Photon number expectation values and variances obtained from mfq- and fq-RT-NEO dynamics with a squeezed coherent state cavity mode initial condition. (a) Photon number expectation value $\nav$. (b) Photon number variance $\varn$.
    (c) mfq-RT-NEO \textit{X}-quadrature variance $\varxx$. 
    (d) mfq-RT-NEO \textit{Y}-quadrature variance $\varyx$.
    (e) fq-RT-NEO \textit{X}-quadrature variance $\varxx$. 
    (f) fq-RT-NEO \textit{Y}-quadrature variance $\varyx$.
    }
    \label{squeeze_2}
\end{figure}

We can extend our analysis of the Mandel Q Parameter calculated with the mfq- and fq-RT-NEO methods by considering the photon number expectation value as well as the photon number and quadrature variances calculated with both methods. 
Figure \ref{squeeze_2}a shows the photon number expectation value $\nav$ calculated with the mfq- and fq-RT-NEO methods. 
The mfq-RT-NEO method predicts very weak fast oscillations with period $\tc/2$ modulated by slow oscillations of $\nav$ with period $\tr$. 
The fq-RT-NEO method, on the other hand, displays fast oscillations of $\nav$ with much greater amplitude but still period $\tc/2$ modulated by slow oscillations with period $\te$.
A similar picture emerges when we consider the photon number variance $\varn$ from the two methods, as shown in Figure \ref{squeeze_2}b. 
Once again, we observe very weak fast oscillations with period $\tc/2$ modulated by slow oscillations with period $\tr$ in the mfq-RT-NEO result. 
This behavior is again contrasted by the fq-RT-NEO result, which shows fast oscillations of $\varn$  with much greater amplitude but still period $\tc/2$ modulated by slow oscillations with period $\te$.
As was found for the corresponding results with a coherent state initial condition, the fast oscillations of $\nav$ and $\varn$ are in phase with those of $\s$, suggesting that the strongly mixed states of the cavity mode have greater photon number expectation value and photon number variance than the weakly mixed states. 

We also consider the quadrature variances $\varxx$ and $\varyx$. 
Figures \ref{squeeze_2}c and \ref{squeeze_2}d show the oscillations of $\varxx$ and $\varyx$, respectively, calculated with the mfq-RT-NEO method. 
Both variances oscillate harmonically with the same amplitude and period $\tc/2$ and are approximately 180 degrees out of phase, indicating that an increase in one quadrature variance corresponds to a decrease in the other quadrature variance by the same magnitude. 
The oscillations in $\varxx$ and $\varyx$ represent a qualitative difference from the corresponding mfq-RT-NEO result for a coherent state initial condition, where these variances remain constant in time. 
The fq-RT-NEO results for the quadrature variances, shown in Figures \ref{squeeze_2}e and \ref{squeeze_2}f, parallel the fq-RT-NEO results obtained with a coherent state initial condition. 
Once again, we observe fast oscillations with period $\tc/2$ modulated by slow Rabi oscillations with period $\te$. 
The Rabi oscillations are in phase between the two variances, while the fast oscillations are exactly 180 degrees out of phase, indicating that an increase in one quadrature variance corresponds to a decrease in the other variance by a comparable magnitude. 
The Rabi oscillations correspond to a modulation of the intensity of the  quadrature squeezing at different times.
They occur in the fq-RT-NEO calculations but are absent in the mfq-RT-NEO results, allowing us to identify them as a consequence of light--matter entanglement between the cavity mode and the molecular vibrational mode. 
As observed for a coherent state initial condition, we also observe that the fast oscillations of $\varxx$ are in phase with those of $\s$.
This observation indicates that the weakly mixed states are \textit{X}-quadrature squeezed, while the strongly mixed states are \textit{Y}-quadrature squeezed.
This result matches our findings with a coherent state initial condition.

We summarize the key dynamics observed in this section as follows. 
The Mandel Q Parameter $\mandelqx$ calculated with the mfq-RT-NEO method and a squeezed coherent state initial condition shows fast oscillations with period $\tc/2$ and slow oscillations with period $\tr$.
This result differs from the result calculated with a coherent state initial condition, where $\mandelqx$ is constant in time.
Moreover, these mfq-RT-NEO dynamics are fundamentally different from those calculated with the fq-RT-NEO method for a squeezed coherent state initial condition. 
Based on the fq-RT-NEO calculations, we can again identify two classes of states between which the cavity mode state oscillates, as we did for a coherent state initial condition. 
Fast maxima of $\mandelqx$ correspond to weakly mixed states with super-Poissonian photon statistics, while fast minima of $\mandelqx$ correspond to strongly mixed states with mostly super-Poissonian photon statistics but also Poissonian statistics occurring with a period of $\te$. 
The weakly mixed states are \textit{X}-quadrature squeezed, while the strongly mixed states are \textit{Y}-quadrature squeezed and have greater photon number expectation values and variances. 
Finally, the fq-RT-NEO dynamics of the quadrature variances show fast harmonic oscillations modulated by slow Rabi oscillations.  
These slow oscillations are absent in the dynamics of the quadratures calculated with the mfq-RT-NEO method. 
All of these differences can be identified as a consequence of including light--matter entanglement in the fq-RT-NEO method. 

\subsection{Fock State (Sub-Poissonian) Initial Condition}
Finally, we consider the photon statistics obtained with the fq-RT-NEO method when the cavity mode initial state $\ket{\psi_{\rm F}(0)}$ is prepared as a single-photon Fock state
\begin{equation}
    \ket{\psi_{\rm F}(0)} = \hat{a}^\dagger \ket{0}
    \label{coherent state def}
\end{equation}
The density operator associated with this single-photon Fock state is the pure state density operator $\hat{\rho}(0) = \hat{a}^\dagger \ket{0}\bra{0}\hat{a}$. Here, $\ket{0}$ is the vacuum state of the single cavity mode (i.e., no photon occupation) and the operator $\hat{a}^\dagger$ creates a single photon in the cavity mode. 
From Eq. \ref{Mandel Q def 2}, the initial value of the Mandel Q Parameter is $Q_{k, x}(0) = -1$.
We have shown in Ref. \citenum{welman_initialization_2026} that polariton formation does not occur with this initial condition when the light--matter interaction is treated at the mean-field level with the mfq-RT-NEO method and only occurs with the fq-RT-NEO method. We therefore focus our attention on the full-quantum results.

\begin{figure}[H]
\centering
\includegraphics[scale=0.35]{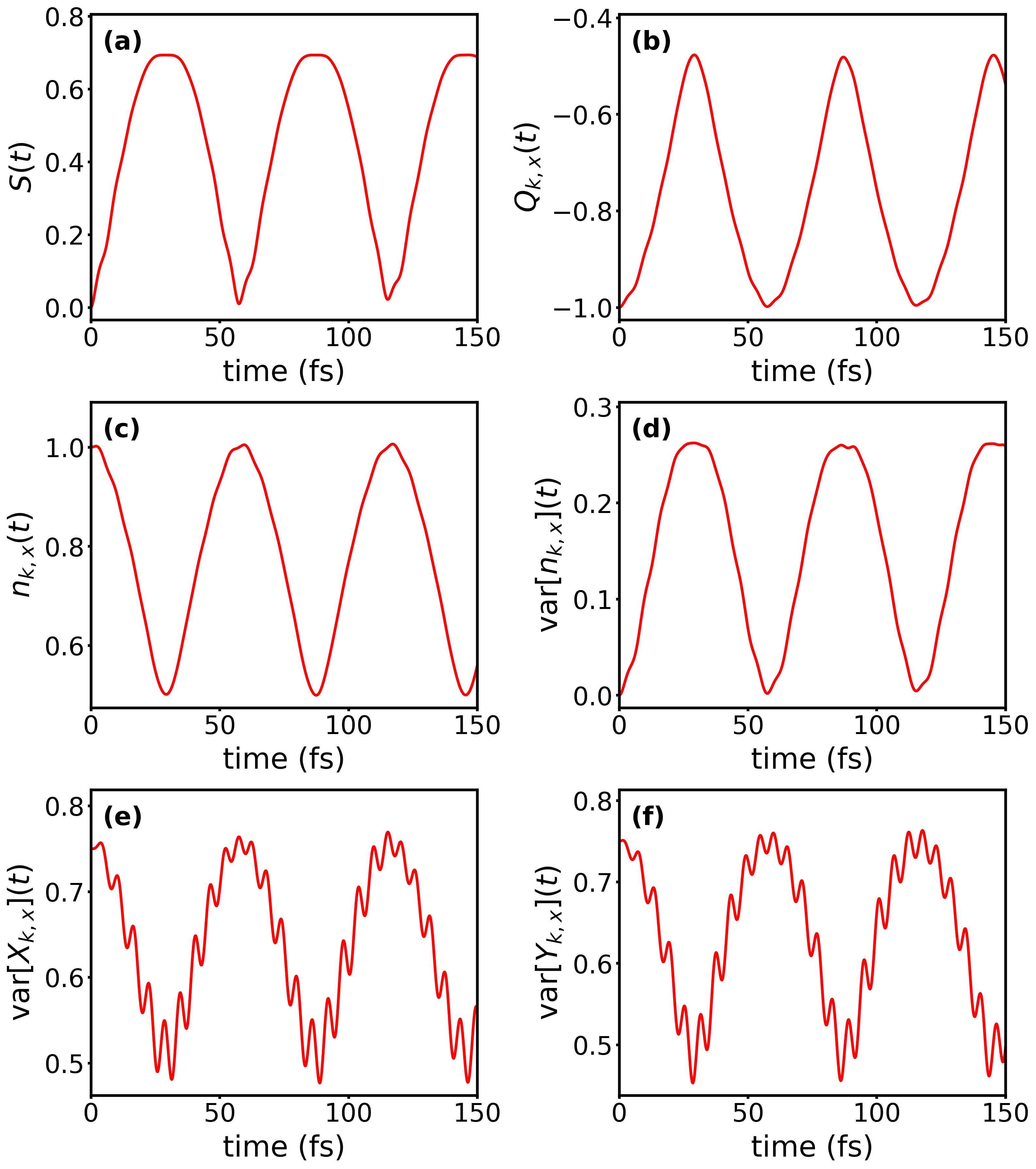}
    \caption{Time-dependent quantities obtained from fq-RT-NEO dynamics with a single-photon Fock state cavity mode initial condition. (a) von Neumann entropy $\s$. (b) Mandel Q Parameter $\mandelqx$. (c) Photon number expectation value $\nav$. (d) Photon number variance $\varn$. 
    (e) \textit{X}-quadrature variance $\varxx$. 
    (f) \textit{Y}-quadrature variance $\varyx$.}
    \label{Fock}
\end{figure}

Figure \ref{Fock}a shows the von Neumann entropy $\s$ calculated with the fq-RT-NEO method. 
Oscillations occur starting from $\s \sim 0.0$ with a period $\te^{\rm Fock} \simeq 59$ fs and a maximum value of $\sim 0.7$.
Note that the period of oscillations of the entropy here, $\te^{\rm Fock}$, is longer than $\te$, the slow oscillation period found with the previous two initial conditions.
The dynamics of the Mandel Q Parameter $\mandelqx$ shown in Figure \ref{Fock}b show oscillations on the same timescale that are in phase with those of $\s$.
This result is different from our results with the coherent state and squeezed coherent state initial conditions, where the oscillations of $\s$ and $\mandelqx$ were exactly 180 degrees out of phase. 
Although the photon statistics are always sub-Poissonian, there is a notable increase toward the Poissonian regime, with a minimum value of $\mandelqx = -1$ rising to a maximum value of $\mandelqx \sim -0.5$. 
To understand the dynamics of $\mandelqx$ better, we can isolate the contributions of the photon number expectation value $\nav$ and photon number variance $\varn$, as shown in Figures \ref{Fock}c and \ref{Fock}d, respectively. 
The oscillations of both $\nav$ and $\varn$ also occur with a period $\te^{\rm Fock}$ and are exactly 180 degrees out of phase with each other. 
This observation indicates that a decrease in the photon number is associated with an increase in the variance, and vice versa. 
Furthermore, the oscillations of $\varn$ are in phase with those of both $\s$ and $\mandelqx$, signifying that increased photon number variance corresponds to a more mixed cavity mode state. 
We can also analyze the quadrature variances $\varxx$ and $\varyx$ shown in Figures \ref{Fock}e and \ref{Fock}f, respectively. Both quadrature variances show slow oscillations with period $\te^{\rm Fock}$ modulating fast oscillations with period $\tc/2$. Although the fast oscillations are 180 degrees out of phase with each other between the two variances, the slow oscillations are perfectly in phase with each other and with the oscillations of $\nav$. 

The phase agreement between the slow oscillations of the quadrature variances illuminates phenomenology that does not occur for either of the two other initial conditions considered in this work. When the cavity mode was initialized either as a coherent state or as a squeezed coherent state, an increase of a given magnitude in one quadrature variance indicated a decrease in the other quadrature variance of around the same magnitude. With a Fock state initial condition, we also observe that a positive fluctuation on short timescales (i.e., $\tc/2$) in one variance is accompanied by a negative fluctuation of the same magnitude in the other variance. Unlike with the other initial conditions, however, we see that on longer timescales (i.e., $\te^{\rm Fock}$), the values of both variances increase and decrease together. A visual description of this phenomenon may be conceived as follows. The quadrature variances can be thought of as defining the major and minor axes of an ellipse in an X-Y plane, centered on the expectation values of the quadrature operators at a given time. For a coherent state or Fock state initial condition, the quadrature variances are equal, so the major and minor axes will be as well, forming a perfect circle. As time progresses, the major and minor axes expand and contract as the quadrature variances increase and decrease. For a coherent state or squeezed coherent state initial condition, corresponding to the fq-RT-NEO dynamics shown in Figures \ref{coherent_2}c and d and Figures \ref{squeeze_2}e and f, respectively, the ellipse will stretch and contract in a manner that returns to its initial shape with period $\tc/2$, while the amplitudes of these stretches and contractions evolve with a longer period $\te$. The situation with a Fock state initial condition, however, is qualitatively different, according to the quadrature variances in Figure \ref{Fock}e and f. Although the major and minor axis lengths fluctuate on a timescale $\tc/2$, on the longer timescale $\te^{\rm Fock}$, the axis lengths of the ellipse uniformly shrink before expanding back to their initial values. In short, the timescale of the ellipse's return to its original shape is much slower for the Fock initial condition ($\te^{\rm Fock}$) than it is for the coherent and squeezed coherent initial conditions ($\tc/2$). Animations of the dynamics of these ellipses in optical phase space are provided in the Supporting Information. 

Taking all of these observations together, we can summarize the cavity mode dynamics with a Fock state initial condition. After initialization in the single-photon Fock state $\ket{1}$, interaction with the molecular vibrational mode causes the cavity mode to evolve into a mixed state with a smaller average photon number $\nav$ but greater photon number variance $\varn$. Because $n_{k, \lambda}(0) = 1$ and ${\rm var}\left[n_{k, \lambda}\right](0) = 0$, these shifts correspond to a smaller difference between $\nav$ and $\varn$ and the cavity mode photon statistics approaching the Poissonian regime. The mixed state into which $\ket{1}$ evolves is also characterized by reduced variance in both of the field quadratures. The cavity mode subsystem returns essentially to the single-photon Fock state $\ket{1}$ with period $\te^{\rm Fock}$. 

\section{Conclusion}
In this work, we have used our mfq- and fq-RT-NEO methods to analyze cavity mode photon statistics under vibrational strong coupling. We considered three different cavity mode initial states: a coherent state, a squeezed coherent state, and a single-photon Fock state. These three initial conditions collectively span Poissonian, super-Poissonian, and sub-Poissonian initial photon statistics, respectively. Comparison of the results of the mfq- and fq-RT-NEO methods allowed us to identify the contribution of light--matter quantum entanglement to the dynamics of the photon statistics. 

For a coherent state initial condition, the fq-RT-NEO method revealed new phenomenology determined to arise from light--matter entanglement because it was absent in the mfq-RT-NEO results. We observed sizable fluctuations of the Mandel Q Parameter away from its initial value, and the cavity mode state was found to rapidly oscillate between weakly mixed and strongly mixed states. The weakly mixed states are characterized by lower photon number expectation value and variance than the strongly mixed states and are \textit{X}-quadrature squeezed, whereas the strongly mixed states are \textit{Y}-quadrature squeezed.
This last result provided a demonstration of light squeezing due to quantum entanglement between an optical cavity mode and a molecular vibrational mode. To the best of our knowledge, such a result has not been simulated previously. 
The quantum fluctuations observed here are particularly notable for a quasi-classical coherent state initial condition.
These basic patterns were also found in the fq-RT-NEO results with a squeezed coherent state initial condition. 
Once again, they could be identified as a consequence of light--matter entanglement because they were absent in the corresponding mfq-RT-NEO results. 

Finally, the use of a Fock state initial condition with the fq-RT-NEO method showed qualitative differences in the dynamics of the photon statistics between this initial condition and the two other initial conditions considered in this work.
The mfq-RT-NEO method is not able to predict this result because, as we have established previously, it does not even predict polariton formation with a Fock state initial condition. Therefore, all three initial conditions considered in this work showed dynamics in the fq-RT-NEO results that could not be obtained by the mfq-RT-NEO method. This finding reflects the role of light--matter entanglement in developing an accurate understanding of the photon statistics. 

These findings suggest that for single-molecule systems under strong coupling to weak cavity fields, consideration of light--matter entanglement may yield insight into new dynamics. 
Weak cavity fields are of low enough intensity that the quantum nature of individual photons may become relevant. 
It is particularly interesting to ask how the effects predicted here would diminish in the limit of large photon number, where the quantum nature of individual photons becomes less important.
In a similar vein, we can also ask how these effects would diminish in the limit of a large number of molecules coupled to the cavity mode. 
These limits correspond to the transition from quantum to classical optics and the transition from the single molecule to the collective coupling regime, respectively. 
Understanding the formal connection between these regimes and our single-molecule quantum optical treatment could help enhance our current understanding of molecular cavity quantum electrodynamics and shed light on basic physical principles of relevance to cavity-modified chemistry. 

\begin{acknowledgement}

This material is based upon work supported by the Air Force Office of Scientific Research under AFOSR Award No. FA9550-24-1-0347. We thank Joel Yuen-Zhou for helpful discussions. 

\end{acknowledgement}

\begin{suppinfo}
Representative Q-Chem input file and molecular geometry for the reported results; animations of optical phase space ellipses.

\end{suppinfo}

\section*{Data and Software Availability}
The data that support the findings of this study are available in the published article,
the Supporting Information, and Zenodo.
\bibliography{Refs}
\end{document}


	\setcounter{equation}{0}
	\setcounter{figure}{0}
	\setcounter{table}{0}
	\renewcommand{\theequation}{S\arabic{equation}}
	\renewcommand{\thefigure}{S\arabic{figure}}
	\renewcommand{\bibnumfmt}[1]{[S#1]}
	\renewcommand{\citenumfont}[1]{S#1}
	\renewcommand{\thepage}{S\arabic{page}}
	
	\maketitle
	
	\newpage


\newpage

\section{Q-Chem Input File for HCN with fq-BO-RT-NEO}
	\RecustomVerbatimCommand{\VerbatimInput}{VerbatimInput}%
	{fontsize=\footnotesize,
		frame=lines,  
		framesep=2em, 
		%
		%
		commandchars=\|\(\), 
		commentchar=*        
	}
	
\begin{Verbatim}
$molecule
0 1
C 0.0 0.0 -0.5026771429
N 0.0 0.0  0.6555628571
H 0.0 0.0 -1.5728771429 
$end

$rem
sym_ignore = 1
input_bohr = false
method = b3lyp
neo = true
neo_epc = epc172
basis = cc-pvdz
SCF_ALGORITHM = GDM
thresh = 14
s2thresh = 12
SCF_CONVERGENCE = 11
NEO_N_SCF_CONVERGENCE = 11
MAX_SCF_CYCLES = 500
NEO_PURECART = 1111
NEO_E_CONV = 12
MEM_TOTAL = 7000
NEO_VPP = 0
NEO_BASIS_LIN_DEP_THRESH = 8 
$end

$neo_basis
H    3
S    1    1.000000
   2.828400 1.0
S    1    1.000000
   4.0 1.0
S    1    1.000000
   5.6569 1.0
S    1    1.000000
   8.0 1.0
S    1    1.000000
   11.3137 1.0
S    1    1.000000
   16.0 1.0
S    1    1.000000
   22.6274 1.0
S    1    1.000000
   32.0 1.0
P    1    1.000000
   2.828400 1.0
P    1    1.000000
   4.0 1.0
P    1    1.000000
   5.6569 1.0
P    1    1.000000
   8.0 1.0
P    1    1.000000
   11.3137 1.0
P    1    1.000000
   16.0 1.0
P    1    1.000000
   22.6274 1.0
P    1    1.000000
   32.0 1.0
D    1    1.000000
   2.828400 1.0
D    1    1.000000
   4.0 1.0
D    1    1.000000
   5.6569 1.0
D    1    1.000000
   8.0 1.0
D    1    1.000000
   11.3137 1.0
D    1    1.000000
   16.0 1.0
D    1    1.000000
   22.6274 1.0
D    1    1.000000
   32.0 1.0
****
$end

$neo_tdks
dt 2.0
photon_type 2
is_coherent true 0
is_squeeze false 0.0 0.0
photon_basis_size 12
qme false
scf_e true
use_RK4 false
maxiter 5000
field_type delta
field_amp 0.3
in_cavity true 0.3475 0 8e-4 1e8
in_cavity_excite_molecule false 
rt_thresh 4
$end
\end{Verbatim}

The input file given above generates the fq-BO-RT-NEO results with a coherent state initial condition shown in Figures 2 and 3 of the main text.
The \textit{\$neo\_tdks} section controls the RT-NEO dynamics. ``\textit{photon\_type}" determines the treatment of the cavity mode. 
It is set to 1 for mean-field quantum calculations and 2 for full-quantum calculations. 
Setting ``\textit{is\_coherent}" to true initializes the cavity mode in a coherent state.
To initialize the cavity mode in a number (Fock) state, ``\textit{is\_coherent}" is set to false, and the subsequent integer determines the number of photons excited in the cavity mode.
Setting ``\textit{is\_squeeze}" to true squeezes the number or coherent state prepared using the previous line by a complex parameter $\xi$ whose real part is given by the first double precision number and whose complex part is given by the second double precision number.
``\textit{scf\_e}" controls the use of the electronic Born--Oppenheimer approximation. ``\textit{in\_cavity true 0.3475 0 8e-4 1e8}" couples the molecule to a single cavity mode with frequency 0.3475 eV, polarization direction $x$ (0 for $x$, 1 for $y$, 2 for $z$), light--matter coupling $\varepsilon = 8 \times 10\textsuperscript{-4}$ a.u., and cavity lifetime $1/\gamma_{\rm c} = 1 \times 10^8$ a.u. Cavity loss is neglible in all of the simulations in this work due to the large cavity lifetime. 
Finally, ``\textit{rt\_thresh}" sets the threshold of the real-time predictor-corrector algorithm to 10\textsuperscript{-4}; this is the \textit{eps} parameter in Algorithm 1 of Ref. \citenum{de_santis_pyberthart_2020}. 
All other parameters are assumed to be self-explanatory.

\bibliography{SI}